\documentclass[prd,showpacs,floatfix,amsmath,nofootinbib,amssymb,floatfix,twocolumn]{revtex4} 
\usepackage{graphicx,color,dcolumn,booktabs,bm}
\usepackage{longtable,lscape}
\usepackage{subfigure}
\usepackage{txfonts}
\usepackage{overpic}
\usepackage{amssymb,amsmath}
\usepackage{indentfirst}
\usepackage{feynmf}   
\usepackage{slashed}  
\usepackage{cases}
\usepackage{color}
\usepackage{multirow}
\usepackage{epstopdf}
\usepackage{longtable}
\usepackage{graphicx,color,dcolumn,booktabs,bm}
\usepackage[colorlinks,
            citecolor=blue,
            anchorcolor=red,
            menucolor=red,
            linkcolor=red,
            filecolor=red,
            runcolor=red,
            urlcolor=blue,
            frenchlinks=red]{hyperref}

\graphicspath{{Figures/}} %

\allowdisplaybreaks

\begin{document}

\title{The bound and resonant states of $D^{(*)}D^{(*)}$ and $D^{(*)}\bar{D}^{(*)}$ with the complex scaling method}

\author{Jia-Liang Lu$^{1}$}
\author{Mao Song$^{1}$}
\email{songmao@ahu.edu.cn}
\author{Peng Wang$^{1}$}
\author{Jian-You Guo$^{1}$}
\author{Gang Li$^{1}$}
\author{Xuan Luo$^{1}$}

\affiliation{$^1$School of Physics and Optoelectronic engineering, Anhui University, Hefei 230601, China}

\begin{abstract}
In this paper, a systematic investigation is conducted to explore the potential molecular states formed by heavy meson pairs $D^{(*)}D^{(*)}$ and $D^{(*)}\bar{D}^{(*)}$ with the complex scaling method, in the framework of the one-boson-exchange (OBE) model. The exchange interactions are mediated by the pseudoscalar, scalar and vector mesons($\pi$, $\sigma$, $\rho$, $\omega$). The interaction potential within the OBE model is derived using the Bonn approximation, followed by the application of the complex scaling method to determine the bound and resonant states. The results demonstrate that the $D^{(*)}D^{(*)}$ and $D^{(*)}\bar{D}^{(*)}$ systems can form not only multiple bound states, but also several P-wave resonant states. In the hadronic molecular state framework, the $X(3872)$, $T_{cc}^+$, and $Z_c(3900)$ states can be consistently explained as bound states, while the $G(3900)$ can be interpreted as a P-wave resonant state. Furthermore, we also predict other new bound and resonant states, which have the potential to be observed experimentally.
\end{abstract}

\pacs{12.39.Pn, 14.40.Lb,25.70.Ef, 25.80.-e}

\maketitle

\section{introduction}\label{sec1}

Most hadrons can be classified into baryons ($qqq$) and mesons ($q\bar{q}$). However, Quantum Chromodynamics (QCD), the fundamental theory of strong interactions, also permits the existence of more complex structures known as exotic states~\cite{Gell-Mann:1964}. Exotic states can generally be classified into two categories: compact multiquark states (e.g., tetraquarks $qq\bar{q}\bar{q}$ and pentaquarks $qqq\bar{q}\bar{q}$) and weakly-bound hadronic molecules composed of two or more conventional hadrons. Since the discovery of $X(3872)$~\cite{Belle:2003nnu,CDF:2003cab,D0:2004zmu,BaBar:2004oro}, numerous exotic hadrons, such as $X$, $Y$, $Z$, and $P_c$ states, have been observed. Exploring their structures and interactions remains a key focus in hadron physics.

\par
To date, three exotic hadron states $X(3872)$, $Z_c(3900)$, and $T_{cc}^+$ have been discovered experimentally, with masses near the $DD^*$ threshold. Various phenomenological models, such as the chiral effective field theory~\cite{Xu:2017tsr,Ohkoda:2012hv,Li:2012cs,Ren:2021dsi}, Bethe-Salpeter approach~\cite{Sakai:2017avl,He:2014nya,He:2015mja,Wallbott:2019dng}, constituent quark models~\cite{Zhu:2019iwm,Ortega:2021yis,Tan:2020ldi,Luo:2017eub}, QCD sum rules~\cite{Navarra:2007yw,Xin:2021wcr,Tang:2019nwv}, and relativized quark models~\cite{Lu:2020rog,Ebert:2007rn,Wang:2018pwi}, have been proposed to describe these states. These models, particularly the hadronic molecule and tetraquark interpretations, provide clear physical insights into the structure of exotic states. Given their proximity to the $DD^*$ threshold, the hadronic molecule explanation is especially natural. If the hadronic molecule model is valid, two hadrons can form not only bound states but also resonant states with higher angular momentum. The study of the resonant states offers valuable insights into the structure of hadronic molecules and the interactions between hadrons.

Resonance behavior is ubiquitous in physical phenomena, appearing in atoms, molecules, nuclei, and chemical reactions. Traditional scattering theory methods, such as the \rm{R}-matrix~\cite{Wigner:1947zz, Hale:1987zz}, \rm{K}-matrix~\cite{Humblet:1991zz}, \rm{J}-matrix~\cite{Taylor}, scattering phase shift, and continuous spectrum theory, are widely used. Additionally, bound-state-like methods, including the real stabilization method (RSM)~\cite{Hazi}, analytic continuation of the coupling constant (ACCC)~\cite{Kukulin}, and complex scaling method (CSM)~\cite{csm1,csm2}, have also been developed. The CSM provides a unified framework for describing bound states, resonant states, and continuum states, making it widely applicable to resonance studies in atomic, molecular, and nuclear physics. Based on its advantages, we extended the CSM to hadronic systems and found that the $DD(\bar{D})$, $\Lambda_c D(\bar{D})$ and $\Lambda_c\Lambda_c(\bar{\Lambda}_c)$ systems can form both bound states and high-angular-momentum resonant states~\cite{Yu:2021lmb}. When applied to Y(4630), the CSM successfully identifies it as a resonant state of the $\Lambda_c \bar{\Lambda}_c$ system. Recently, the CSM has been widely adopted in hadron physics~\cite{Wang:2022yes,Cheng:2022qcm,Wang:2023ivd,Lin:2024prl,Lin:2022wmj,Chen:2023eri}.

In this work, we will employ the CSM to systematically investigate bound and resonant states near the heavy mesons $D^{(*)}D^{(*)}$ and $D^{(*)}\bar{D}^{(*)}$ threshold. The $D^{(*)}D^{(*)}$ and $D^{(*)}\bar{D}^{(*)}$ S-wave bound states have been extensively studied within the OBE model~\cite{Li:2012cs,Li:2012ss,Liu:2019stu,Abreu:2022sra,Abreu:2015jma}. However, P-wave resonant states remain understudied and have only recently attracted research interest~\cite{Lin:2024prl,Whyte:2024ihh}. As we know, the $X(3872)$ is widely interpreted as a loose bound state of $D\bar{D}^{*}$ with quantum number $J^{PC} = 1^{++}$ ~\cite{Wang:2013kva,Thomas:2008ja,Braaten:2010mg}. The ratio $\mathcal{B}[X(3872)\to J/\psi \pi^+\pi^-\pi^0]/\mathcal{B}[X(3872)\to J/\psi \pi^+\pi^-]$ indicates significant isospin breaking in the hidden-charm decay of $X(3872)$ \cite{D0:2004zmu,Belle:2005lfc,BaBar:2010wfc}. $T^+_{cc}(cc\bar{u}\bar{d})$ is considered as molecular structure of $DD^{*}$ with quantum number $I(J^P) = 0(1^+)$~\cite{Sakai:2023syt,Ohkoda:2012hv,Cheng:2022qcm,Ren:2021dsi,Liu:2019stu}. Although, the interaction of $I=1$ is weaker than that of $I=0$ in $D\bar{D}^*$ systems, the $Z_c(3900)$ was also suggested to be an isovector $D\bar{D}^*$ molecule with quantum numbers $J^{PC}=1^{+-}$~\cite{Guo:2013sya,Wang:2013cya}. Recent theoretical studies suggest that $G(3900)$ can be interpreted as a P-wave resonant state of the $D\bar{D}^*/\bar{D}D^*$~\cite{Lin:2024prl,Huang:2025rvj,Chen:2025gxe,Liu:2025sjz}. However, the scattering amplitudes global analysis for the processes $e^+e^- \to D\bar{D},D\bar{D}^*+c.c.$, and $D^*\bar{D}^*$ indicates the $G(3900)$ as a dynamically generated state~\cite{Ye:2025ywy}. Whether the molecular state explanations of these exotic hadron states are reasonable and whether they can form other bound and resonant states still need to be further studied both in theory and experiments.

This paper is organized as follows. After the introduction, Section \ref{sec2} provides the theoretical framework and calculation methods. Section \ref{sec3} presents the numerical results and discussion, followed by a summary in Section \ref{sec4}.

\section{Theoretical framework}\label{sec2}
The dynamics of the $\mathrm{P}^{(\ast)}\mathrm{P}^{(\ast)}$ hadronic molecule respects two key symmetries: heavy quark symmetry and chiral symmetry. Chiral perturbation theory (CPT) serves as the low-energy effective field theory of QCD. In this theoretical framework, the interaction between heavy mesons can be described by the interaction term between chiral fields and heavy meson fields. This interaction Lagrangian is invariant under the heavy quark spin transformation and chiral transformation~\cite{Nambu:1961tp,Burdman:1992gh,Wise:1992hn,Yan:1992gz,Casalbuoni:1996pg,Manohar:2000dt,Isola:2003fh}. By appropriately choosing coupling constants and interaction terms in the Lagrangian, the interactions of heavy mesons under chiral symmetry can be described. Chiral symmetry and its spontaneous breaking play a crucial role in heavy hadron systems.

To derive the $\mathrm{P}^{(\ast)}\mathrm{P}^{(\ast)}$ potential, effective Lagrangians are introduced to describe interactions between heavy mesons mediated by the exchange of the pseudoscalar meson $\pi$, vector mesons ($v = \rho ,\omega$) and scalar meson $\sigma$~\cite{Ding:2008gr,Sakai:2023syt,Ohkoda:2012hv}. The interaction Lagrangians are given as

\begin{eqnarray}
 {\cal L}_{\pi \rm{PP}^*} &=&
 - \frac{g}{f_\pi}(\rm{P}^\dagger_a \rm{P}^\ast_{b\,\mu}+\rm{P}^{\ast\,\dagger}_{a\,\mu}\rm{P}_b)\partial^\mu\hat{\pi}_{ba}
 \, ,
 \label{eq:piPP*}\\
 {\cal L}_{\pi \rm{P}^*\rm{P}^*} &=&
  i \frac{g}{f_\pi}\epsilon^{\mu \nu \alpha \beta } v_{\mu}
 \rm{P}^{\ast\,\dagger}_{a\,\nu}\rm{P}^\ast_{b\,\alpha}\partial_{\beta}
 \hat{\pi}_{ba}
\label{eq:piP*P*} \, , \\
{\cal L}_{v\rm{PP}} &=&
\sqrt{2}\beta g_V \rm{P}_b \rm{P}^{\dagger}_a v\cdot \hat{\rho}_{ba}
 \, ,
\label{eq:vPP} \\
{\cal L}_{v\rm{PP}^*} &=&
-2\sqrt{2}\lambda g_V \epsilon^{\mu \nu \alpha \beta}v_{\mu}
\left(\rm{P}^{\ast\,\dagger}_{a\,\nu}\rm{P}_b+\rm{P}^{\dagger}_a\rm{P}^{\ast}_{b\,\nu}\right)
\partial_\alpha(\hat{\rho}_\beta)_{ba}
\, ,
\label{eq:vPP*} \\
{\cal L}_{v\rm{P}^*\rm{P}^*} &=&
\sqrt{2} \beta g_V \rm{P}^*_b \rm{P}^{*\dagger}_a v \cdot \hat{\rho}_{ba}\nonumber \\
&&+i2\sqrt{2} \lambda g_V
\rm{P}^{\ast\,\dagger}_{b\,\mu}\rm{P}^\ast_{a\,\nu}
(\partial^\mu(\hat{\rho}^\nu)_{ba}-\partial^\nu(\hat{\rho}^\mu)_{ba}) \, ,
\label{eq:vP*P*} \\
{\cal L}_{\sigma \rm{PP}} &=& - 2 g_s \rm{P}^{\dagger} \rm{P} \sigma,
\label{eq:sPP1} \\
{\cal L}_{\sigma \rm{P}^*\rm{P}^*} &=& 2 g_s \rm{P}^{*\dagger}_\mu \rm{P}^{*\mu} \sigma.
\label{eq:sPP2}
\end{eqnarray}
The interaction term ${\cal L}_{\pi \rm{P}\rm{P}} = 0$ due to the parity conservation, where $\rm{P} = (\rm{D}^0, \rm{D}^+)$ and $\rm{P}^* = (\rm{D}^{*0}, \rm{D}^{*+})$. The subscripts $a$ and $b$ denote light flavor indices (up and down), and $v_{\mu}$ is a four-velocity fixed as $v_{\mu}=(1,\vec{0})$ in the following. The pseudoscalar meson $\pi$ and vector meson $\rho$ fields are defined as follows:

\begin{eqnarray}
\hat{\pi} &=& \left(
\begin{array}{cc}
 \frac{\pi^{0}}{\sqrt{2}} & \pi^{+} \\
 \pi^{-} & -\frac{\pi^{0}}{\sqrt{2}}
\end{array}
\right)= \frac{\vec{\tau}\cdot\vec{\pi}}{\sqrt{2}},  \\
  \hat{\rho}_{\mu} &=&  \displaystyle{\left(
\begin{array}{cc}
 \frac{\rho^{0}}{\sqrt{2}} + \frac{\omega}{\sqrt{2}}& \rho^{+} \\
 \rho^{-} & -\frac{\rho^{0}}{\sqrt{2}} + \frac{\omega}{\sqrt{2} }
\end{array}
\right)}_{\mu} \, = \frac{\vec{\tau}\cdot\vec{\rho}_{\mu}}{\sqrt{2}}.
\end{eqnarray}

Here, $g$, $\lambda$, $\beta$ and $g_V$ are coupling constants in the
interaction Lagrangians, while $f_{\pi}$ and $m_{\rho}$ denote the pion decay constant and $\rho$ meson mass, respectively.
The one-pion exchange potentials (OPEPs) are derived from the interaction Lagrangians (\ref{eq:piPP*}) and (\ref{eq:piP*P*}) as follows:
\begin{eqnarray}
V^{\pi}_{P_{1}P_{2}^{\ast} \rightarrow P_{1}^{\ast}P_{2}} &\!=\!& \left( \frac{g}{2 f_{\pi}} \right)^{2} \frac{1}{3} [ -\vec{\varepsilon}_{1}^{\ast} \cdot \vec{\varepsilon}_{2} D(r;m_{\pi})+\vec{\varepsilon}_{1}^{\ast} \cdot \vec{\varepsilon}_{2}  C(r;m_{\pi})  \nonumber \\
&& +S_{\varepsilon_{1}^{\ast},\varepsilon_{2}} \, T(r;m_{\pi}) ]\vec{\tau}_{1}  \cdot \vec{\tau}_{2},  \nonumber \\
V^{\pi}_{P_{1}P_{2}^{\ast} \rightarrow P_{1}^{\ast}P_{2}} &\!=\!& \left( \frac{g}{2 f_{\pi}} \right)^{2} \frac{1}{3} [ -\vec{T}_{1} \!\cdot\! \vec{T}_{2} D(r;m_{\pi})+\vec{T}_{1} \!\cdot\! \vec{T}_{2}  C(r;m_{\pi}) \nonumber \\
&& +S_{T_{1},T_{2}} \, T(r;m_{\pi}) ]\vec{\tau}_{1}  \cdot \vec{\tau}_{2},  \label{eq:pot_P*P*P*P*} \nonumber \\
\end{eqnarray}
where $m_{\pi}$ is the $\pi$ meson mass. Here polarization vectors for $\mathrm{P}^{*}$ are defined as
$\vec{\varepsilon}^{\hspace{0.2em}(\pm)} \!=\! \left(\mp 1/\sqrt{2}, - i/\sqrt{2}, 0 \right)$ and
$\vec{\varepsilon}^{\hspace{0.2em}(0)} \!=\! \left(0, 0, 1\right)$,
and the spin-1 operator $\vec{T}$ is defined as $T_{\lambda' \lambda}^{i}=i \varepsilon^{ijk} \varepsilon_{j}^{(\lambda')\dag} \varepsilon_{k}^{(\lambda)}$.
By convention, $\vec{\varepsilon}^{\,(\lambda)}$ and $\vec{\varepsilon}^{\,(\lambda)\ast}$ denote incoming and outgoing vector particles, respectively. Here $\vec{\tau}_{1}$ and $\vec{\tau}_{2}$ are isospin operators for $\mathrm{P}^{(\ast)}_{1}$ and
$\mathrm{P}^{(\ast)}_{2}$, with $\vec{\tau}_{1}  \!\cdot\! \vec{\tau}_{2} = -3$ for total isospin $\rm{I}=0$ and 1 for $\rm{I}=1$. The tensor operators are defined as,
\begin{eqnarray}
S_{\varepsilon_{1}^{\ast},\varepsilon_{2}} &=& 3 ( \vec{\varepsilon}^{\,(\lambda_{1})\ast} \!\cdot\!\hat{r} ) ( \vec{\varepsilon}^{\,(\lambda_{2})} \!\cdot\!\hat{r} ) -  \vec{\varepsilon}^{\,(\lambda_{1})\ast} \!\cdot\! \vec{\varepsilon}^{\,(\lambda_{2})}, \nonumber \\
S_{T_{1},T_{2}} &=& 3 ( \vec{T}_{1} \!\cdot\!\hat{r} ) ( \vec{T}_{2} \!\cdot\!\hat{r} ) - \vec{T}_{1} \!\cdot\! \vec{T}_{2},
\end{eqnarray}
Here, $\hat{r}=\vec{r}/r$ is the unit vector between the two mesons.

The $\rho$ and $\sigma$ exchange potentials are derived similarly from the interaction Lagrangians(\ref{eq:vPP})-(\ref{eq:sPP2}),
\begin{eqnarray}
V^{\rho}_{P_{1}P_{2} \rightarrow P_{1}P_{2}} &\!=\!&
 \left( \frac{\beta g_V}{2m_{\rho}} \right)^{2}  C(r;m_{\rho})
 \vec{\tau}_{1}  \!\cdot\! \vec{\tau}_{2}, \label{eq:rhopot_BBBB} \nonumber \\
V^{\rho}_{P_{1}P_{2}^{\ast} \rightarrow
P_{1}P_{2}^{\ast}} &\!=\!&
 \left( \frac{\beta g_V}{2m_{\rho}} \right)^{2}  C(r;m_{\rho})  \vec{\tau}_{1}  \!\cdot\! \vec{\tau}_{2}, \label{eq:rhopot_BB*BB*}  \nonumber \\
 V^{\rho}_{P_{1}P_{2}^{\ast} \rightarrow
  P_{1}^{\ast}P_{2}} &\!=\!&
 \left( \lambda g_V \right)^{2} \frac{1}{3} [-2\vec{\varepsilon}_{1}^{\ast} \cdot \vec{\varepsilon}_{2} D(r;m_{\pi})+
  2\vec{\varepsilon}_{1}^{\,\ast} \!\cdot\! \vec{\varepsilon}_{2} \,
  C(r;m_{\rho}) \nonumber \\
 &&\!-\! S_{\varepsilon_{1}^{\ast},\varepsilon_{2}} \, T(r;m_{\rho})
					     ]  \vec{\tau}_{1}
 \!\cdot\! \vec{\tau}_{2}, \label{eq:rhopot_BB*B*B} \nonumber \\
V^{\rho}_{P_{1}^{\ast}P_{2}^{\ast} \rightarrow P_{1}^{\ast}P_{2}^{\ast}} &\!=\!&
 \left(  \lambda g_V \right)^{2} \frac{1}{3} [ -2\vec{T}_{1} \!\cdot\! \vec{T}_{2} D(r;m_{\pi}) + 2\vec{T}_{1}
  \!\cdot\! \vec{T}_{2} \, C(r;m_{\rho}) \nonumber  \\
 &&\!-\! S_{T_{1},T_{2}} \,
  T(r;m_{\rho}) ]  \vec{\tau}_{1}  \!\cdot\! \vec{\tau}_{2}  + \left( \frac{\beta g_V}{2m_{\rho}} \right)^{2} C(r;m_{\rho})
 \vec{\tau}_{1}  \!\cdot\! \vec{\tau}_{2},  \label{eq:rhopot_B*B*B*B*} \nonumber \\
V^{\sigma}_{P_{1}P_{2} \rightarrow P_{1}P_{2}} &\!=\!&
- \left( \frac{ g_s}{m_{\sigma}} \right)^{2} C(r;m_{\sigma}), \label{eq:sigma_BBBB} \nonumber \\
V^{\sigma}_{P_{1}P_{2}^{\ast} \rightarrow P_{1}P_{2}^{\ast}} &\!=\!&
- \left( \frac{ g_s}{m_{\sigma}} \right)^{2} \vec{\varepsilon}_{2}^{\,\ast} \!\cdot\! \vec{\varepsilon}_{4} C(r;m_{\sigma}), \label{eq:sigma2_BBBB} \nonumber \\
V^{\sigma}_{P_{1}^{\ast}P_{2}^{\ast} \rightarrow P_{1}^{\ast}P_{2}^{\ast}} &\!=\!&
- \left(\frac{ g_s}{m_{\sigma}} \right)^{2} \vec{\varepsilon}_{1}^{\,\ast} \!\cdot\! \vec{\varepsilon}_{3}\vec{\varepsilon}_{2}^{\,\ast} \!\cdot\! \vec{\varepsilon}_{4}C(r;m_{\sigma}). \label{eq:sigma3_BBBB}
\end{eqnarray}

Here, the $D(r)$ term represents the delta function, and arises from short-range interactions, which lie beyond the valid range of the OBE interaction. Two alternative approaches are commonly adopted for the $D(r)$ term: retaining it~\cite{Wang:2022mxy,Liu:2009qhy,Wang:2020dya,Chen:2021tip,Wang:2024ukc} or removing it~\cite{Thomas:2008ja,Liu:2019zvb,Ling:2021asz}. Recently, the authors introduce a parameter to phenomenologically adjust the strength of the $D(r)$ term, systematically examining how the $D(r)$ potential affects the near threshold states~\cite{Xu:2025mhc,Yalikun:2025ssz}. The optimal treatment of the $D(r)$ term remains an open question that requires further investigation. Following the Ref.~\cite{Sakai:2023syt}, we have discarded the $D(r)$ term. The $\omega$ meson exchange potentials are obtained by replacing the $\rho$ meson mass with $\omega$ meson mass and removing the isospin factor $\vec{\tau}_{1}  \!\cdot\! \vec{\tau}_{2}$. The OPEP's for $\mathrm{P}^{(\ast)}\bar{\mathrm{P}}^{(\ast)}$ differ from those for $\mathrm{P}^{(\ast)}\mathrm{P}^{(\ast)}$ by an overall sign change due to $G$-parity, and the same applies to the $\omega$ meson
exchange potentials. However, the $\rho$ and $\sigma$ meson exchange potentials for $\mathrm{P}^{(\ast)}\bar{\mathrm{P}}^{(\ast)}$ remain unchanged due to their even $G$-parity~\cite{Ohkoda:2011vj,Ding:2008gr}.

In the above equations, $C(r;m_{h})$ and $T(r;m_{h})$  are defined as follows:
\begin{eqnarray}
\hspace{-3em}&&C(r;m) \!=\! \int \frac{\mbox{d}^{3}\vec{q}}{(2\pi)^3} \frac{m^{2}}{\vec{q}^{\,\,2}+m^{2}}
 e^{i\vec{q} \cdot \vec{r}} \,
\mathcal{F}(\vec{q};m), \\
\hspace{-3em}&&T(r;m) S_{12}(\hat{r}) \!=\! \int \frac{\mbox{d}^{3}\vec{q}}{(2\pi)^3} \frac{- \vec{q}^{\,\,2}}{\vec{q}^{\,\,2}+m^{2}}
S_{12}(\hat{q})e^{i\vec{q} \cdot \vec{r}} \mathcal{F}(\vec{q};m),
\end{eqnarray}
with $S_{12}(\hat{r}) \!=\! 3 (\vec{\mathcal{O}}_{1} \!\cdot\! \hat{r})
(\vec{\mathcal{O}}_{2} \!\cdot\! \hat{r}) - \vec{\mathcal{O}}_{1} \!\cdot\!
\vec{\mathcal{O}}_{2}$.

The form factor is introduced to suppress the high-momentum contributions in meson-exchange interactions since the light mesons couple to heavy mesons as composite systems rather than probing their internal structure. The form factor usually adopts monopole, dipole, or exponential forms. At low momenta ($q^2 \ll \Lambda^2$), these forms give similar results if we adjust $\Lambda$ value, since their first-order Taylor expansions coincide. This means hadronic molecule properties at low energy don't change much when using different form factors. The dipole form factor provides stronger high-momentum suppression, and the exponential form factor features Gaussian-type momentum suppression and is particularly suitable for heavy quarkonium. The monopole form factor has the simplest functional form and is widely used in hadronic molecular state. Here, a monopole form factor $\mathcal{F}(\vec{q};m)$ is introduced at each vertex, defined as follows:
\begin{eqnarray}
\mathcal{F}(\vec{q};m)\!=\! \frac{\Lambda^{2} \!-\!
m^{2}}{\Lambda^{2} \!+\! \vec{q}^{\,\,2}}  \, ,
\end{eqnarray}
Here, $\Lambda$ is the cutoff parameter, while $m$ and $\vec{q}$ denote the mass and momentum of the exchanged meson ($= \pi, \rho, \omega, \sigma$), respectively. In Refs~\cite{Yasui:2009bz,Yamaguchi:2011xb,Ohkoda:2011vj}, $\Lambda$ is related to the root-mean-square (RMS) radius of the source hadron. After Fourier transformation, the central and tensor functions are obtained:
\begin{align}
  C(r;m) = & \frac{m^2}{4\pi}\left[
    \frac{e^{-mr}}{r} - \frac{e^{-\Lambda r}}{r} - \frac{\Lambda^2-m^2}{2\Lambda}e^{-\Lambda r}
  \right]\label{eq;central_fun_explicit},\\
  T(r;m) = &
  \frac{1}{4\pi}(3 + 3mr + m^2r^2
 )\frac{e^{-mr}}{r^3}
  -\frac{1}{4\pi}(3 + 3\Lambda r + \Lambda^2r^2)\frac{e^{-\Lambda r}}{r^3}\notag\\
  &+\frac{1}{4\pi}\frac{m^2 - \Lambda^2}{2}(1+\Lambda r)\frac{e^{-\Lambda r}}{r}.
\end{align}

Using the meson-exchange potentials $\mathcal{V}(r)$ in coordinate space, we can solve the non-relativistic Schr\"{o}dinger equation to obtain the bound state eigenvalues and eigenfunctions. For the resonant state, the CSM is adopted by introducing an unbounded and nonunitary operator $U(\theta)$ with a rotation angle $\theta$. The basic idea of the complex scaling method is to perform a coordinate transformation by replacing the real coordinate $r$ with complex coordinate $re^{i\theta}$. By applying the coordinate transformation to the Schr\"{o}dinger equation, the Hamiltonian of the system is modified. The modified Hamiltonian, denoted as $H_{\theta}$, is obtained by replacing $H_\theta(r) =U(\theta)H(r)U(\theta)^{-1}$. According to the Aguilar-Balslev-Combes theorem~\cite{abc}, the resonant solutions of the the complex scaled Schr\"{o}dinger equation are square integrable when the rotation angle $\theta > \theta_c$, where $\theta_c$ is the critical angle separating the resonant state from the continuum. We solve the complex-scaled Schr\"{o}dinger equation using the basis expansion method, with the radial function expressed in spherical harmonic oscillator bases. For the detailed calculation scheme, please refer to our previous work.~\cite{Yu:2021lmb}.

\section{Numerical results}\label{sec3}

In this section, we will present the numerical results and discussion. The relevant parameters and the meson masses are listed in Table~\ref{parameter}.
Following the Ref.\cite{Sakai:2023syt}, the coupling $\sigma$ meson constant $g_s$ is estimated by one-third of the coupling strength between nucleon and $\sigma$ meson.
By diagonalizing the modified Hamiltonian, we obtain the eigenvalues and eigenvectors of the bound and resonant states.

\begin{table}[!htbp]
  \caption{The relevant parameters are used in this work \cite{Sakai:2023syt}.}\label{parameter}
  \begin{tabular}{ccc|cc}\toprule[2pt]
  Hadron       &$I(J^P)$     &Mass (MeV)    &Parameters   & \\\hline
  $\pi$        &$1(0^-)$                    &138          &g   &  0.59  \\
  $\rho$       &$1(1^-)$          &770          &$g_V$     & $\frac{m_\rho}{\sqrt{2}f_\pi}$   \\
  $\omega$     &$0(1^-)$          &782          &$\beta$   & 0.9   \\
  $\sigma$     &$0(0^+)$          &500          &$\lambda$ & 0.56 $\rm{GeV^{-1}}$     \\
  $\mathrm{D}$        &$\frac{1}{2}(0^-)$   &1868         &$g_s$     & 3.4    \\
  $\mathrm{D}^{\ast}$  &$\frac{1}{2}(1^-)$  &2009         &$f_{\pi}$  &  93 MeV   \\
  \bottomrule[2pt]
  \end{tabular}
\end{table}

The pseudoscalar meson $D$ and vector meson $D^{\ast}$ can form three combinations: $DD$, $DD^{\ast}$, and $D^{\ast}D^{\ast}$. These states are classified by isospin $I$, total angular momentum $J$, and parity $P$. Table \ref{num2} lists the states with quantum numbers $I^G(J^{PC})$ and their corresponding wave function channels. Note that the wave functions must be symmetric under the exchange of two $D^{(\ast)}$ mesons. In this case, the total wave function for the combined systems of $D^{(\ast)}$ and $D^{(\ast)}$ must be symmetric under group $O(3)$ $\times$ $SU_I(2)$ $\times$ $SU_S(2)$, where $SU_I(2)$ and $SU_S(2)$ are isospin and spin groups respectively. The isosinglet and isovector wave functions are antisymmetric and symmetric, respectively. In order to obtain symmetric wave functions, $\frac{1}{\sqrt{2}}(D^*D \pm DD^*)$ structures are introduced~\cite{Ohkoda:2012hv,Ke:2021rxd}. The narrow near-threshold peaks observed in $D^0D^0$ and $D^+D^0$ mass spectra support the $T_{cc}^+$ decay mechanism via intermediate $D^*$ meson formation, followed by $D^* \to D\pi/D\gamma$ decays~\cite{LHCb:2021vvq,LHCb:2021auc}. Based on experimental observations, interpreting the $T_{cc}^+$ as an isoscalar $0(1^+)$ hadronic molecular state is physically reasonable. However, the $T_{cc}^+$ wave function is dominated by the $D^{*+}D^0$ component, because its mass is very close to the $D^{*+}D^0$ threshold~\cite{Du:2021zzh}.

For the $D^{(\ast)}\bar{D}^{(\ast)}$ system, $C$-parity must be considered. The states are classified by the quantum numbers $I^G(J^{PC})$, where where $I$, $G$, $J$, $P$, and $C$ denote isospin, $G$-parity, total angular momentum, parity, and charge conjugation, respectively. These states are listed in Table \ref{num3}. The charge conjugation $C$ is defined for $I=0$ or $I_Z=0$ components of $I=1$, with $G=(-1)^I C$. For $I=0$, many $D^{(\ast)}\bar{D}^{(\ast)}$ states share the same $J^{PC}$ quantum numbers as quarkonia, as indicated in Table \ref{num3}. As is known, tensor forces play a crucial role in bound state formation, thus angular momentum mixing is considered. In order to reduce uncertainties and highlight the dominant molecular components, we have neglected coupled-channel effects in this paper. As shown in Ref.~\cite{Lin:2024prl,Liu:2025fhl}, the results demonstrate that the coupled-channel effects are non-negligible in these processes, but do not alter the formation of the bound and resonant states.

\begin{table*}[htp]
  \caption{\small Possible $D^{(\ast)}D^{(\ast)}(^{2S+1}L_{J})$ channels for given quantum numbers $I$ and $J^{P}$, with $J \le 2$. }\label{num2}
  \begin{tabular}{|c|c|c|c|c|}
\hline
$I$ & $J^{P}$ & $DD$ & $DD^{\ast}$&$D^{\ast}D^{\ast}$ \\
\hline
 & $0^{-}$ & &$\frac{1}{\sqrt{2}}( DD^{\ast} + D^{\ast}D )(^{3}P_{0})$ &\\
\cline{2-5}
 & $1^{+}$ & & $\frac{1}{\sqrt{2}} \left( DD^{\ast}-D^{\ast}D \right) (^{3}S_{1})$, $\frac{1}{\sqrt{2}} \left( DD^{\ast}-D^{\ast}D \right) (^{3}D_{1})$ & $D^{\ast}D^{\ast}(^{3}S_{1})$, $D^{\ast}D^{\ast}(^{3}D_{1})$ \\
\cline{2-5}
0 & $1^{-}$ & $DD(^{1}P_{1})$ &$\frac{1}{\sqrt{2}} \left( DD^{\ast}+D^{\ast}D \right)(^{3}P_{1})$ & $D^{\ast}D^{\ast}(^{1}P_{1})$, $D^{\ast}D^{\ast}(^{5}P_{1})$, $D^{\ast}D^{\ast}(^{5}F_{1})$ \\
\cline{2-5}
 & $2^{+}$ & & $\frac{1}{\sqrt{2}} \left( DD^{\ast}-D^{\ast}D \right)(^{3}D_{2})$ & $D^{\ast}D^{\ast}(^{3}D_{2})$  \\
\cline{2-5}
 & $2^{-}$ & &$\frac{1}{\sqrt{2}} \left( DD^{\ast}+D^{\ast}D \right)(^{3}P_{2})$, $\frac{1}{\sqrt{2}} \left( DD^{\ast}+D^{\ast}D \right)(^{3}F_{2})$ & $D^{\ast}D^{\ast}(^{5}P_{2})$, $D^{\ast}D^{\ast}(^{5}F_{2})$\\
\hline
 & $0^{+}$ &$DD(^{1}S_{0})$ & & $D^{\ast}D^{\ast}(^{1}S_{0})$, $D^{\ast}D^{\ast}(^{5}D_{0})$\\
 \cline{2-5}
 & $0^{-}$ & & $\frac{1}{\sqrt{2}} \left( DD^{\ast}-D^{\ast}D \right)(^{3}P_{0})$ & $D^{\ast}D^{\ast}(^{3}P_{0})$\\
 \cline{2-5}
1 & $1^{+}$ & &$\frac{1}{\sqrt{2}} \left( DD^{\ast}+D^{\ast}D \right) (^{3}S_{1})$, $\frac{1}{\sqrt{2}} \left( DD^{\ast}+D^{\ast}D \right)(^{3}D_{1})$ & $D^{\ast}D^{\ast}(^{5}D_{1})$\\
\cline{2-5}
 & $1^{-}$ & &$\frac{1}{\sqrt{2}} \left( DD^{\ast}-D^{\ast}D \right)(^{3}P_{1})$ & $D^{\ast}D^{\ast}(^{3}P_{1})$\\
 \cline{2-5}
 & $2^{+}$ & $DD(^{1}D_{2})$ & $\frac{1}{\sqrt{2}} \left( DD^{\ast}+D^{\ast}D \right)(^{3}D_{2})$ & $D^{\ast}D^{\ast}(^{1}D_{2})$, $D^{\ast}D^{\ast}(^{5}S_{2})$, $D^{\ast}D^{\ast}(^{5}D_{2})$, $D^{\ast}D^{\ast}(^{5}G_{2})$\\
\cline{2-5}
 & $2^{-}$ & &$\frac{1}{\sqrt{2}} \left( DD^{\ast}-D^{\ast}D \right)(^{3}P_{2})$, $\frac{1}{\sqrt{2}} \left( DD^{\ast}-D^{\ast}D \right)(^{3}F_{2})$ & $D^{\ast}D^{\ast}(^{3}P_{2})$, $D^{\ast}D^{\ast}(^{3}F_{2})$ \\
\hline
  \end{tabular}
\end{table*}

\begin{table*}[htp]
  \caption{\small Possible $D^{(\ast)}\bar{D}^{(\ast)}(^{2S+1}L_{J})$ channels for given quantum numbers $I$ and $J^{PC}$, with $J \le 2$. Exotic quantum numbers that cannot be attributed to conventional charmonium states ($c\bar{c}$) are explicitly marked with the symbol $\surd$. }\label{num3}
  \begin{tabular}{|c|c|c|c|c|c|}
\hline
  $J^{PC}$ & $D\bar{D}$ & $D\bar{D}^{\ast}/\bar{D}D^{\ast}$ &$D^{\ast}\bar{D}^{\ast}$ &$I=0$ &$I=1$\\
\hline
  $0^{++}$ &$D\bar{D}(^{1}S_{0})$ & & $D^{\ast}\bar{D}^{\ast}(^{1}S_{0})$, $D^{\ast}\bar{D}^{\ast}(^{5}D_{0})$ &$\chi_{c0}$& $\surd$\\
 \cline{1-6}
 $0^{--}$ & &$\frac{1}{\sqrt{2}}( D\bar{D}^{\ast} + D^{\ast}\bar{D} )(^{3}P_{0})$ & &$\surd$ & $\surd$\\
\cline{1-6}
 $0^{-+}$ & & $\frac{1}{\sqrt{2}} \left( D\bar{D}^{\ast}-D^{\ast}\bar{D} \right)(^{3}P_{0})$ & $D^{\ast}\bar{D}^{\ast}(^{3}P_{0})$ &$\eta_{c}$ &$\surd$\\
  \cline{1-6}
  $1^{+-}$ & & $\frac{1}{\sqrt{2}} \left( D\bar{D}^{\ast}-D^{\ast}\bar{D} \right) (^{3}S_{1})$, $\frac{1}{\sqrt{2}} \left( D\bar{D}^{\ast}-D^{\ast}\bar{D} \right) (^{3}D_{1})$ & $D^{\ast}\bar{D}^{\ast}(^{3}S_{1})$, $D^{\ast}\bar{D}^{\ast}(^{3}D_{1})$ &$h_{c}$&$\surd$\\
\cline{1-6}
  $1^{++}$ & &$\frac{1}{\sqrt{2}} \left( D\bar{D}^{\ast}+D^{\ast}\bar{D} \right) (^{3}S_{1})$, $\frac{1}{\sqrt{2}} \left( D\bar{D}^{\ast}+D^{\ast}\bar{D} \right)(^{3}D_{1})$ & $D^{\ast}\bar{D}^{\ast}(^{5}D_{1})$&$\chi_{c1}$ &$\surd$\\
\cline{1-6}
  $1^{--}$ & $D\bar{D}(^{1}P_{1})$ &$\frac{1}{\sqrt{2}} \left( D\bar{D}^{\ast}+D^{\ast}\bar{D} \right)(^{3}P_{1})$ & $D^{\ast}\bar{D}^{\ast}(^{1}P_{1})$, $D^{\ast}\bar{D}^{\ast}(^{5}P_{1})$, $D^{\ast}\bar{D}^{\ast}(^{5}F_{1})$&$J/\psi$ &$\surd$\\
  \cline{1-6}
    $1^{-+}$ & &$\frac{1}{\sqrt{2}} \left( D\bar{D}^{\ast}-D^{\ast}\bar{D} \right)(^{3}P_{1})$ & $D^{\ast}\bar{D}^{\ast}(^{3}P_{1})$ &$\surd$ & $\surd$\\
 \cline{1-6}
  $2^{+-}$ & & $\frac{1}{\sqrt{2}} \left( D\bar{D}^{\ast}-D^{\ast}\bar{D} \right)(^{3}D_{2})$ & $D^{\ast}\bar{D}^{\ast}(^{3}D_{2})$  &$\surd$ & $\surd$ \\
\cline{1-6}
  $2^{++}$ & $D\bar{D}(^{1}D_{2})$ & $\frac{1}{\sqrt{2}} \left( D\bar{D}^{\ast}+D^{\ast}\bar{D} \right)(^{3}D_{2})$ & $D^{\ast}\bar{D}^{\ast}(^{1}D_{2})$, $D^{\ast}\bar{D}^{\ast}(^{5}S_{2})$, $D^{\ast}\bar{D}^{\ast}(^{5}D_{2})$, $D^{\ast}\bar{D}^{\ast}(^{5}G_{2})$ &$\chi_{c2}$ &$\surd$\\
\cline{1-6}
  $2^{-+}$ & &$\frac{1}{\sqrt{2}} \left( D\bar{D}^{\ast}-D^{\ast}\bar{D} \right)(^{3}P_{2})$, $\frac{1}{\sqrt{2}} \left( D\bar{D}^{\ast}-D^{\ast}\bar{D} \right)(^{3}F_{2})$ & $D^{\ast}\bar{D}^{\ast}(^{3}P_{2})$, $D^{\ast}\bar{D}^{\ast}(^{3}F_{2})$ &$\eta_{c2}$ &$\surd$ \\
\cline{1-6}
 $2^{--}$ & &$\frac{1}{\sqrt{2}} \left( D\bar{D}^{\ast}+D^{\ast}\bar{D} \right)(^{3}P_{2})$, $\frac{1}{\sqrt{2}} \left( D\bar{D}^{\ast}+D^{\ast}\bar{D} \right)(^{3}F_{2})$ & $D^{\ast}\bar{D}^{\ast}(^{5}P_{2})$, $D^{\ast}\bar{D}^{\ast}(^{5}F_{2})$ &$\psi_{2}$&$\surd$\\
\hline
  \end{tabular}
\end{table*}

\begin{figure}
\centering
\includegraphics[width=0.5\textwidth]{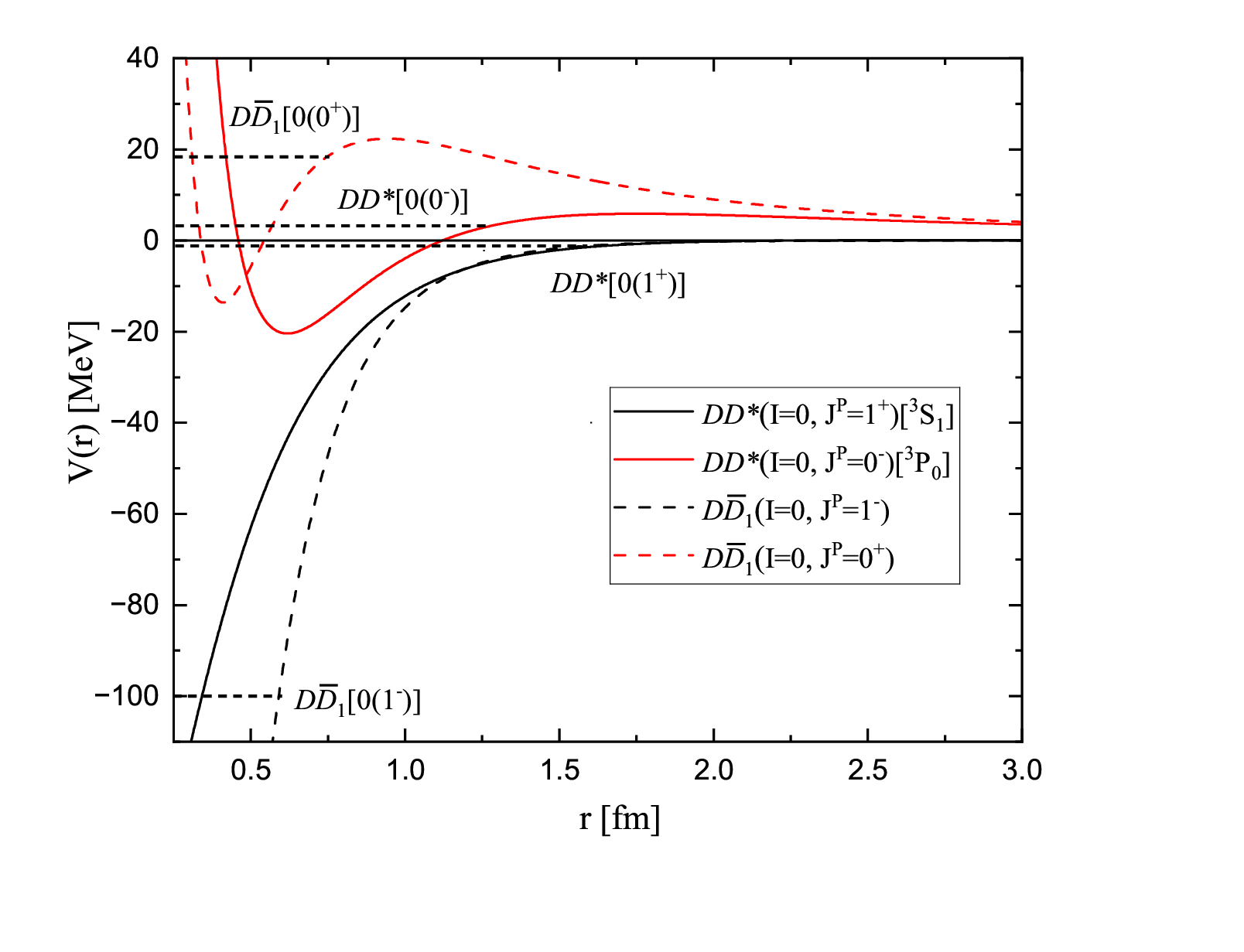}
\vspace*{-1.0cm}
\caption{(Color online) The potential functions of the states $0(1^+)$ and $0(0^-)$ for $D^{(\ast)}D^{(\ast)}$ system. }\label{potential}
\end{figure}

The mass of $T^+_{cc}$ is close to the threshold of $DD^{*}$, with quantum number of $0(1^+)$. We first adjust the cutoff parameter $\Lambda$, and when $\Lambda$ = 1.1426 $\rm{GeV}$, the state $0(1^+)$ of $DD^{*}$ appears a bound state with energy -0.273 $\rm{MeV}$, which is consistent with the binding energy of $T^+_{cc}$. When the cutoff parameter $\Lambda$ = 1.1426 $\rm{GeV}$, we solve the Schr\"{o}dinger equations for all quantum states of the $D^{(\ast)}D^{(\ast)}$ and $D^{(\ast)}\bar{D}^{(\ast)}$ systems using CSM. The Hamiltonian matrices for all states are given in the appendix~\ref{appendix;H}. From Table~\ref{tbl:result_table_DD}, we can see that $DD$ system cannot form any bound or resonant state. In the $DD^{*}$ system, the $ 0(1^+)$ state is identified as a bound state, which is consistent with the properties of the $T^+_{cc}$ tetraquark state. Additionally, the $0(0^-)$ state is found to be a resonant state with a mass excess of 3.86 $\rm{MeV}$ above the $DD^*$ threshold and a decay width of 27.68 $\rm{MeV}$. Compared with the S-wave interaction potential, the P-wave interaction potential involved a centrifugal potential barrier, in the form of $L(L+1)/2 \mu r^2$. Intuitively, the S-wave is shallow bound state, indicating weak interaction between the two mesons, the P-wave resonant state should be more unstable, and the decay width should be larger. However, the S-wave is a deep bound system, and the width of the P-wave resonant state is actually larger for the $D\bar{D}_1$, $D^*\bar{D}_1$ and $D^*\bar{D}^*_2$ systems, as shown in Ref.~\cite{Yu:2021lmb}. In order to clarify the reasons, we present the potentials of S and P-waves for $DD^*$ and $D\bar{D}_1$ systems in the coordinate space, where the P-wave potential is the sum of the one boson exchange potential and the centrifugal potential barrier in Fig.\ref{potential}. We can see that the peak of potential for the P-wave in the $DD^*$ system is lower and the potential barrier is wider, while the peak of the potential for the P-wave in the $D\bar{D}_1$ system is higher and the width of the potential barrier is narrower. This potential barrier allow particles to generate resonant state above the threshold, and the decay width is related to the shape of the potential barrier. The higher the potential barrier, the higher the energy of the resonant state, and the wider the potential barrier, the smaller the width of the resonant state. In addition, the decay width may also be related to the relative velocity of the two mesons. In general, this is an open question, and it deserves further exploration. In other cases, no bound or resonant states have been found. For $D^{\ast}D^{\ast}$ system, there are two bound states with quantum number of $0(1^+)$ and $1(0^+)$.

For $D^{(\ast)}\bar{D}^{(\ast)}$ system with isospin $I=0$, when slightly adjusting the parameter $\Lambda$, we find that at $\Lambda$ = 1.073 $\rm{GeV}$, the $0^+(1^{++})$ state of the $D\bar{D}^{\ast}/D^{\ast}\bar{D}$ system forms a bound state with binding energy -5.031 $\rm{MeV}$, matching the quantum numbers of $X(3872)$. Simultaneously, the isoscalar ($I^G=0^-$) $D\bar{D}^{\ast}/\bar{D}D^{\ast}$ system with $J^{PC}=1^{--}$ forms a resonant state, consistent with both the recently observed $G(3900)$ by BESIII ~\cite{BESIII:2024ths} and the theoretical predictions in Ref.~\cite{Lin:2024prl}. Furthermore, the results reveal that states $D\bar{D}(I^G=0^+,J^{PC}=0^{++})$ and $D\bar{D}^{\ast}/\bar{D}D^{\ast}(I^G=0^-,J^{PC}=1^{+-})$ are both bound states, while we predict two new resonant states $D\bar{D}^{\ast}/\bar{D}D^{\ast}(I^G=0^+,J^{PC}=0^{-+})$ and $D^{\ast}\bar{D}^{\ast}(I^G=0^+,J^{PC}=0^{-+})$. The corresponding results are presented in Table~\ref{tbl:result_table_DDbar2}.

For $D^{(\ast)}\bar{D}^{(\ast)}$ system with isospin $I=1$, the attractive potentials are weaker than that with isospin $I=0$. When the cutoff parameter $\Lambda \geq 2 \rm{GeV}$, the bound states emerge for $D^{(\ast)}\bar{D}^{(\ast)}$ system with isospin $I=1$. Here, we set the cutoff parameter $\Lambda$ as 1.25 $\rm{GeV}$ for $D^{(\ast)}\bar{D}^{(\ast)}$ system with isospin $I=1$. In Table~\ref{tbl:result_table_DDbar}, we present the energies for $D^{(\ast)}\bar{D}^{(\ast)}$ systems with $I=1$. In the $D\bar{D}$ system, no bound or resonant states are found. For the $D\bar{D}^{\ast}/D^{\ast}\bar{D}$ system, two bound states are identified with quantum numbers $1^+(1^{+-})$ and $1^-(1^{++})$. Among these, the $1^-(1^{++})$ state is a bound state with a binding energy of -0.357 $\rm{MeV}$. The $1^+(1^{+-})$ state is a bound state, with a binding energy of -0.085 $\rm{MeV}$, which quantum number matches $Z_c(3900)$. For $D^{\ast}\bar{D}^{\ast}$ system, there are two bound states, which are $1^-(0^{++})$ and $1^+(1^{+-})$ states.

If the interpretation of $X(3872)$, $T_{cc}^+$, and $Z_c$ states as loosely bound states is physically reasonable, then their low-energy observables should be insensitive to the microscopic details of the interactions. The long-range feature of these systems only depends on the scattering length or binding energy. The strong and radiative decay widths can also provide important information about those structure~\cite{Meng:2021jnw}. The existence of the $G(3900)$ is a natural consequence of hadronic molecules like the $X(3872)$, $T_{cc}^+$, and $Z_c$. Unlike bound states, a resonance is a metastable state whose dominant decay channels typically involve its constituent particles (e.g., $DD^*$,$D\bar{D}^{\ast}/\bar{D}D^{\ast}$ and $D^{\ast}\bar{D}^{\ast}$). This provides a unique experimental approach to identify resonance states and investigate the structure of hadronic states~\cite{Lin:2024prl}.

\begin{table}[htp]
\caption{\small The energies of $D^{(\ast)}D^{(\ast)}$ states with quantum numbers $I(J^{P})$ with $J \le 2$. The energies $E$ can be purely real for bound states or complex for resonant states, where the imaginary part corresponds to half of the decay width, $\Gamma/2$. The values in parentheses following the energies represent the root-mean-square (RMS) radii in units of fm. The notation $\ldots$ indicates no bound or resonant state solutions. The blank entries indicate that the states with these quantum numbers are forbidden.}
\begin{center}
{\renewcommand\arraystretch{1.5}
\begin{tabular}{|c|c|c|c|c|}
\hline
$I$ & $J^{P}$ & $DD$ & $DD^{\ast}$ &$D^{\ast}D^{\ast}$ \\
 \cline{3-5}
 & & $\mathrm{E}$ [MeV]& $\mathrm{E}$ [MeV] & $\mathrm{E}$ [MeV]\\
\cline{1-5}
 & $0^{-}$ &  & $3.86-i\frac{27.68}{2}$  &  \\
\cline{2-5}
 & $1^{+}$ &   & -0.273(6.1),$~T^+_{cc}$ & -0.563(4.4) \\
\cline{2-5}
0 & $1^{-}$& \ldots & \ldots  & \ldots\\
\cline{2-5}
 & $2^{+}$ &  &\ldots   & \ldots \\
\cline{2-5}
 & $2^{-}$ &  & \ldots  &  \ldots \\
\hline
 & $0^{+}$ &\ldots  &  & -0.495(4.8) \\
\cline{2-5}
 & $0^{-}$ &  &  \ldots& \ldots \\
\cline{2-5}
1 & $1^{+}$ &  &  \ldots  & \ldots  \\
\cline{2-5}
 & $1^{-}$ &  &  \ldots  & \ldots  \\
\cline{2-5}
 & $2^{+}$ &\ldots &\ldots  & \ldots  \\
\cline{2-5}
 & $2^{-}$ &  &  \ldots  & \ldots  \\
\hline
\end{tabular}
}
\end{center}
\label{tbl:result_table_DD}
\end{table}%

\begin{table}[htp]
\caption{\small The energies of $D^{(\ast)}\bar{D}^{(\ast)}$ states with $I^G(J^{PC})$ in $I = 0$ with $J \le 2$. The convention is the same as in Table~\ref{tbl:result_table_DD}.}
\begin{center}
{\renewcommand\arraystretch{1.5}
\begin{tabular}{|c|c|c|c|}
\hline
$I^G(J^{PC})$ & $D\bar{D}$ & $D\bar{D}^{\ast}$/$\bar{D}D^{\ast}$ &$D^{\ast}\bar{D}^{\ast}$ \\
 \cline{2-4}
 & $\mathrm{E}$ [MeV] & $\mathrm{E}$ [MeV] & $\mathrm{E}$ [MeV]\\
\cline{1-4}
 $0^+(0^{++})$ &-1.658(2.18)  &  & \ldots \\
\cline{1-4}
$0^-(0^{--})$ &   &\ldots  &   \\
\cline{1-4}
 $0^+(0^{-+})$ &  &$3.84-i\frac{27.31}{2}$  & $1.56-i\frac{28.80}{2}$\\
\cline{1-4}
 $0^-(1^{+-})$ &  & -0.370(1.47)  & \ldots \\
 \cline{1-4}
 $0^+(1^{++})$ &  &  -5.031(0.92), $~X(3872)$ &  \ldots \\
  \cline{1-4}
 $0^-(1^{--})$ & \ldots  & $4.25-i\frac{56.62}{2}$, $~G(3900)$  & \ldots  \\
   \cline{1-4}
 $0^+(1^{-+})$ &  & \ldots  & \ldots  \\
\cline{1-4}
 $0^-(2^{+-})$ &  &\ldots & \ldots \\
 \cline{1-4}
 $0^+(2^{++})$ & \ldots &\ldots  & \ldots  \\
  \cline{1-4}
 $0^-(2^{--})$ &  &\ldots  & \ldots \\
   \cline{1-4}
 $0^+(2^{-+})$ &  & \ldots  & \ldots   \\
\hline
\end{tabular}
}
\end{center}
\label{tbl:result_table_DDbar2}
\end{table}%

\begin{table}[htp]
\caption{\small The energies of $D^{(\ast)}\bar{D}^{(\ast)}$ states with $I^G(J^{PC})$ in $I = 1$ with $J \le 2$. The convention is the same as in Table~\ref{tbl:result_table_DD}.}
\begin{center}
{\renewcommand\arraystretch{1.5}
\begin{tabular}{|c|c|c|c|}
\hline
$I^G(J^{PC})$ & $D\bar{D}$ & $D\bar{D}^{\ast}$/$\bar{D}D^{\ast}$ &$D^{\ast}\bar{D}^{\ast}$ \\
 \cline{2-4}
 & $\mathrm{E}$ [MeV] & $\mathrm{E}$ [MeV] & $\mathrm{E}$ [MeV]\\
\cline{1-4}
 $1^-(0^{++})$ &\ldots  &  & -0.167(5.3) \\
\cline{1-4}
$1^+(0^{--})$ &   &\ldots  &   \\
\cline{1-4}
 $1^-(0^{-+})$ &  &\ldots  & \ldots \\
\cline{1-4}
 $1^+(1^{+-})$ &  & -0.085(6.5),$~Z_c(3900)$  & -0.279(4.9) \\
 \cline{1-4}
 $1^-(1^{++})$ &  &  -0.357(4.4) &  \ldots  \\
  \cline{1-4}
 $1^+(1^{--})$ & \ldots  & \ldots  & \ldots  \\
   \cline{1-4}
 $1^-(1^{-+})$ &  & \ldots  & \ldots  \\
\cline{1-4}
 $1^+(2^{+-})$ &  &\ldots & \ldots \\
 \cline{1-4}
 $1^-(2^{++})$ & \ldots &\ldots  & \ldots  \\
  \cline{1-4}
 $1^+(2^{--})$ &  &\ldots  & \ldots \\
   \cline{1-4}
 $1^-(2^{-+})$ &  & \ldots  & \ldots   \\
\hline
\end{tabular}
}
\end{center}
\label{tbl:result_table_DDbar}
\end{table}%

\begin{figure}
\centering
\includegraphics[width=0.5\textwidth]{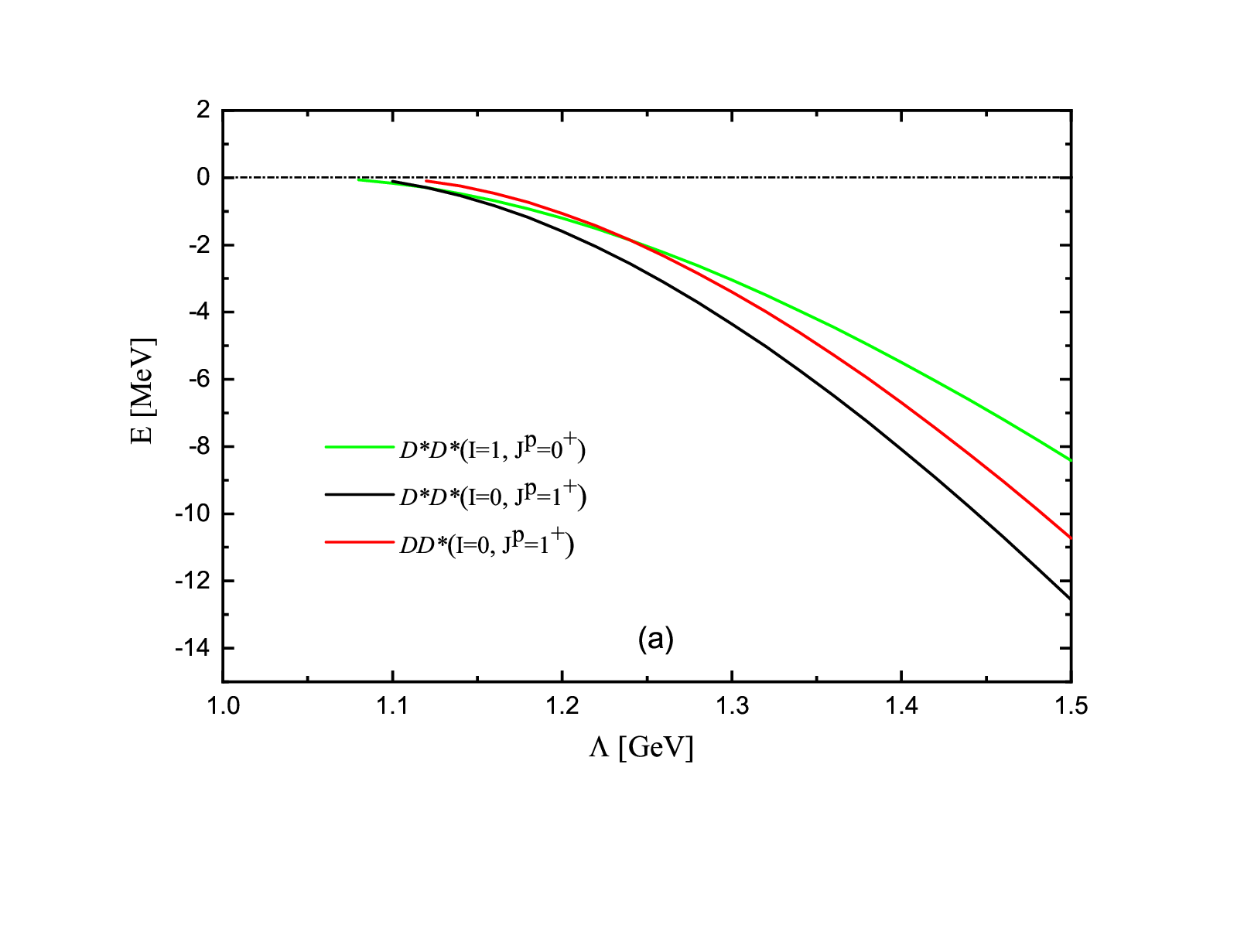}
\includegraphics[width=0.5\textwidth]{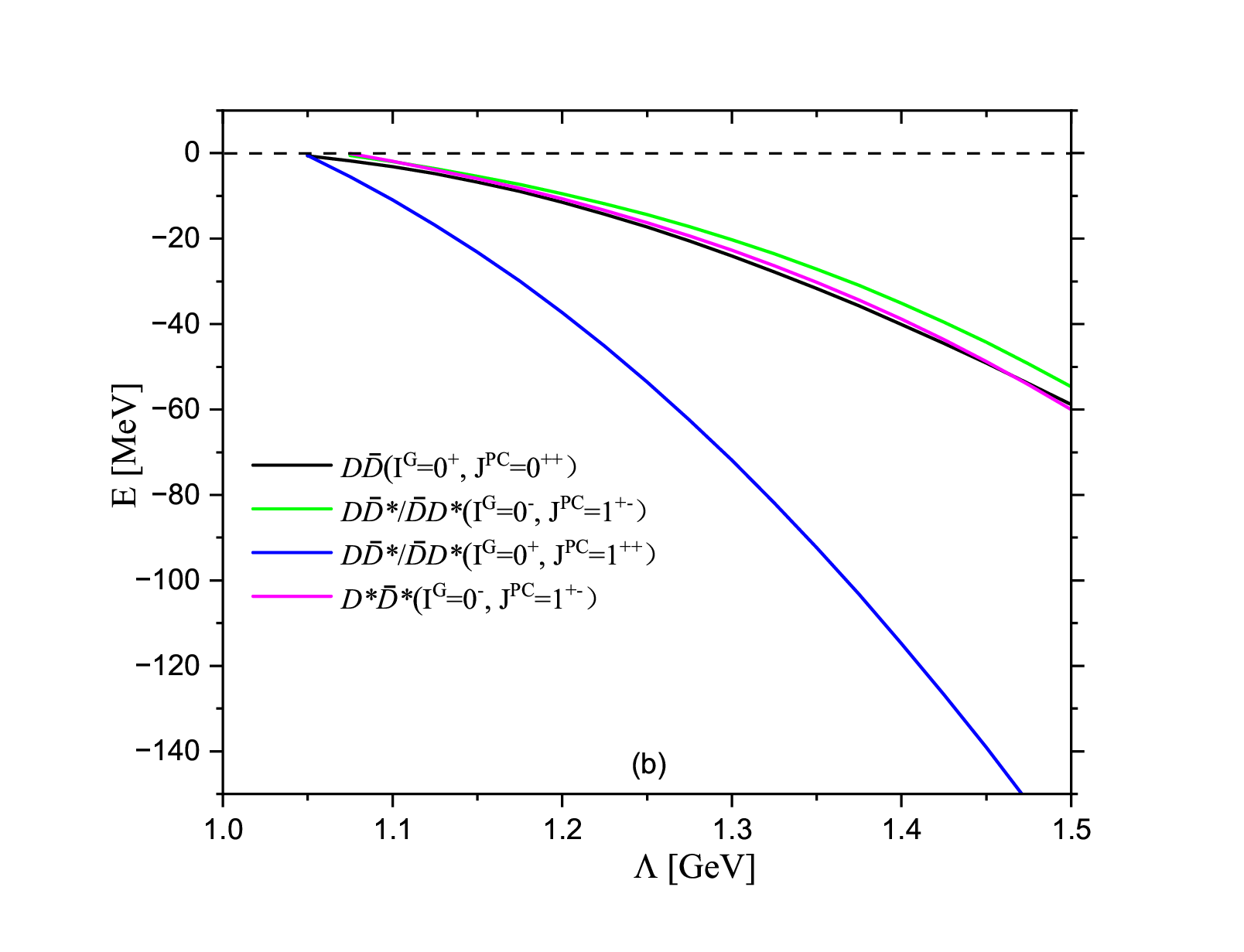}
\includegraphics[width=0.5\textwidth]{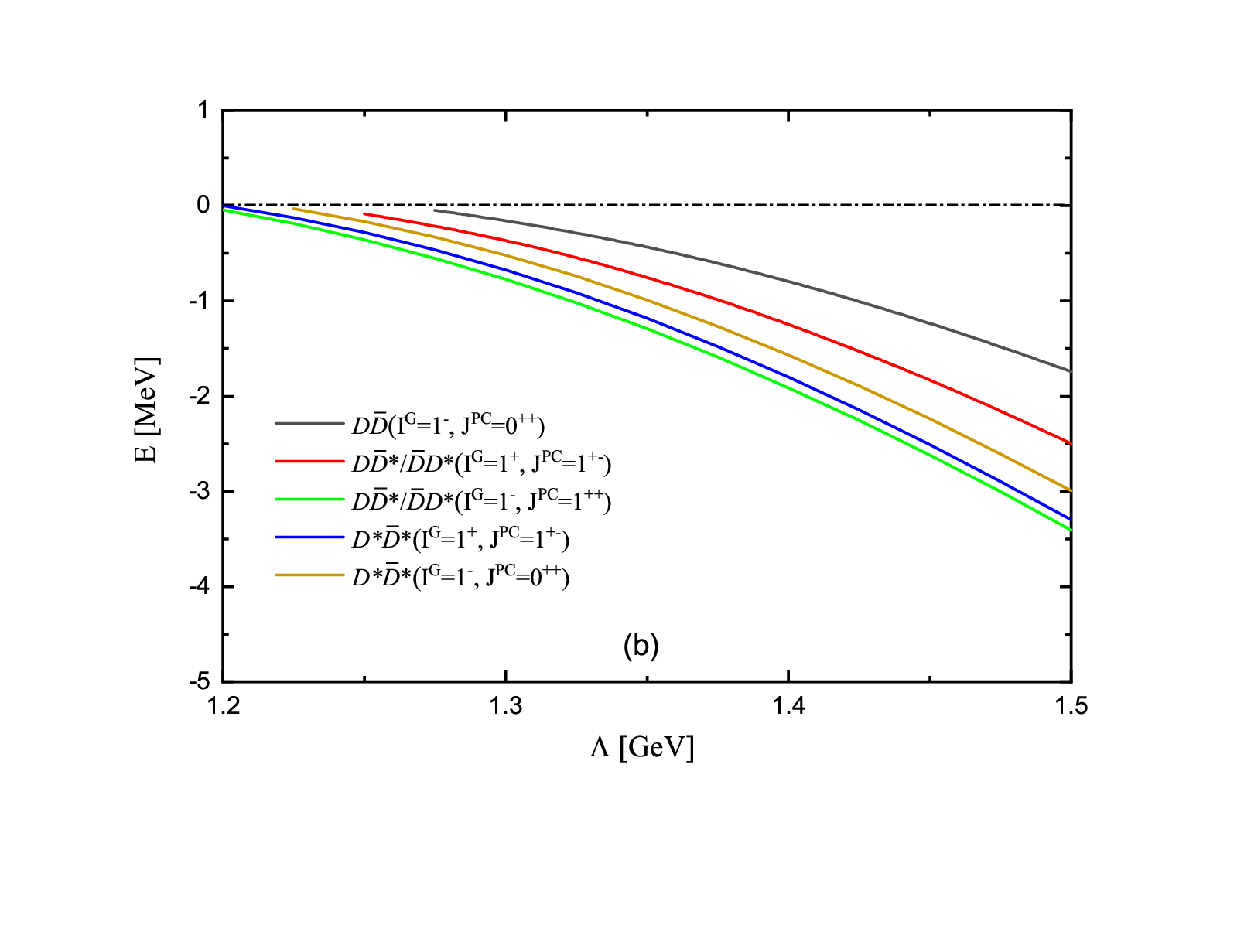}
\vspace*{-0.5cm}
\caption{(Color online) The energies of the bound states as a function of the cutoff parameter $\Lambda$ for $D^{(\ast)}D^{(\ast)}$ and $D^{(\ast)}\bar{D}^{(\ast)}$ systems. }\label{cutoff1}
\end{figure}

\begin{figure}
\centering
\includegraphics[width=10cm]{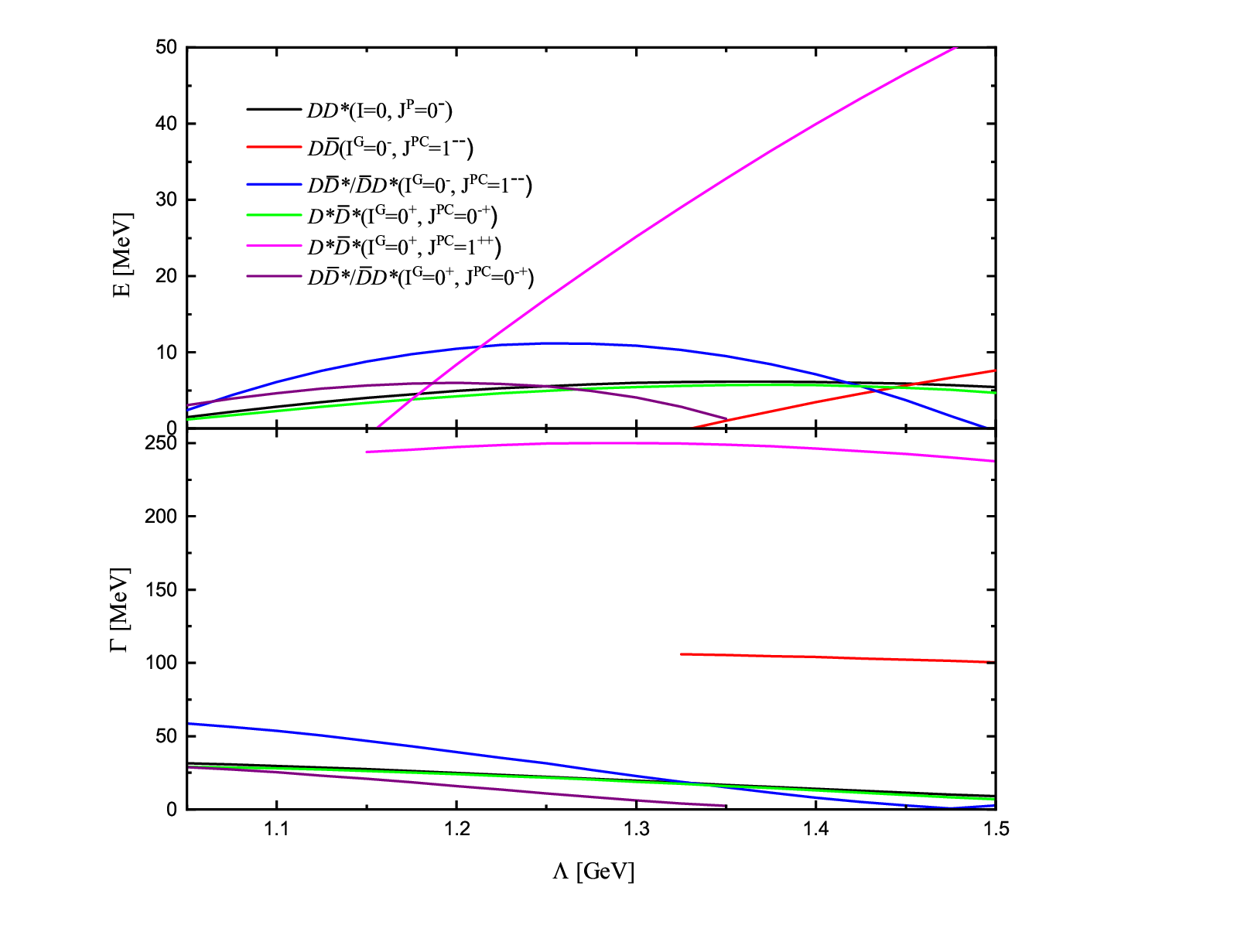}\\
\vspace*{-0.5cm}
\caption{(Color online) The energy and width of the resonant state as a function of the cutoff parameter $\Lambda$ for $D^{(\ast)}D^{(\ast)}$ and $D^{(\ast)}\bar{D}^{(\ast)}$ systems. }\label{resonance}
\end{figure}

The value of cutoff parameter $\Lambda$ is related to the radius of meson, which has a significant impact on the energy and width of the bound and resonant state.
The inverse of the cutoff parameter (1/$\Lambda$) characterizes the typical interaction distance between constituent hadrons in the molecular state. A larger value of $\Lambda$ corresponds to a shorter interaction range, indicating more localized interactions between the components. Conversely, a smaller $\Lambda$ value implies a longer interaction range, reflecting a more diffuse molecular structure where the constituent hadrons are spatially more extended and loosely bound. For nucleon-nucleon interaction, the cutoff parameter $\Lambda$ is usually ranges from 0.8 to 1.5 GeV. In Fig.\ref{cutoff1}, we present the energies of the bound states $D^{(\ast)}D^{(\ast)}$ and $D^{(\ast)}\bar{D}^{(\ast)}$ systems as a function of the cutoff parameter $\Lambda$, respectively. From Fig.\ref{cutoff1}(a), we can see that the isoscalar  $DD^{\ast}(I=0,J^P=1^+)$ and $D^{\ast}{D}^{\ast}(I=0,J^P=1^+)$ are loose bound states. When $\Lambda$ is greater than about 1.1 $\rm{GeV}$, bound states begin to appear, and as $\Lambda$ increases, the binding energies gradually increase. An isovector $J^P=0^+$ bound state appears at $\Lambda\approx1.07~\rm{GeV}$ for $D^{\ast}D^{\ast}$, the binding energy decreases slower than two isoscalar states with the increase of the cutoff parameter $\Lambda$. In Fig.\ref{cutoff1}(b), we provide energy change of the bound states with the cutoff parameter $\Lambda$ for $D^{(\ast)}\bar{D}^{(\ast)}$ system with isospin $I=0$. It can be seen that there are four bound states with isospin $I=0$, and the energies of the bound states are much bigger than those with isospin $I=1$, which range from several $\rm{MeV}$ to dozens of $\rm{MeV}$, and one can reach more than 100 $\rm{MeV}$ for the state $D{\bar{D}^{\ast}}/\bar{D}D^{\ast}(I^G=0^+,J^{PC}=1^{++})$. From Fig.\ref{cutoff1}(c), we can see that the $D^{(\ast)}\bar{D}^{(\ast)}$ system with isospin $I=1$ can form five bound states. Among these bound states, the changing trends of the $D^{(\ast)}\bar{D}^{(\ast)}(I^G=1^+,J^{PC}=1^{+-})$ and $D{\bar{D}^{\ast}}/\bar{D}D^{\ast}(I^G=1^-,J^{PC}=1^{++})$, $D\bar{D}(I^G=1^-,J^{PC}=0^{++})$ and $D{\bar{D}^{\ast}}/\bar{D}D^{\ast}(I^G=1^+,J^{PC}=1^{+-})$ states curves are consistent with the change of cutoff parameter $\Lambda$, because their meson-exchange potentials are similar.

$D^{(\ast)}D^{(\ast)}$ and $D^{(\ast)}\bar{D}^{(\ast)}$ systems can not only form the bound states, but also form several resonant states.
In Fig.\ref{resonance}, we present the energies and widths for the resonant states as a function of cutoff parameter $\Lambda$. We can see that the energy and width changing trends of $D{D}^{\ast}(I=0,J^P=0^-)$ and ${D}^{\ast}\bar{D}^{\ast}(I^G=0^+,J^{PC}=0^{-+})$ states curves are consistent with the change of cutoff parameter $\Lambda$, the energies increases to its maximum value, then slowly decreases with the increasing of cutoff parameter $\Lambda$, the corresponding width slowly decreases with the increasing of cutoff parameter $\Lambda$. The energies of the resonant states are approximately a few $\rm{MeV}$, and the widths vary from a few $\rm{MeV}$ to several tens of $\rm{MeV}$. ${D}^{\ast}\bar{D}^{\ast}(I^G=0^+,J^{PC}=1^{++})$ state is a D-wave resonant state, and its width is the largest. The energy is gradually increasing with the increase of cutoff parameter $\Lambda$, but the width does not change much around 250 $\rm{MeV}$. $D\bar{D}(I^G=0^-,J^{PC}=1^{--})$ state is a P-wave resonant state, and its width is much larger than other P-wave resonant states, which almost unchanged with the cutoff parameter $\Lambda$.

\begin{figure}
\centering
\includegraphics[width=10cm]{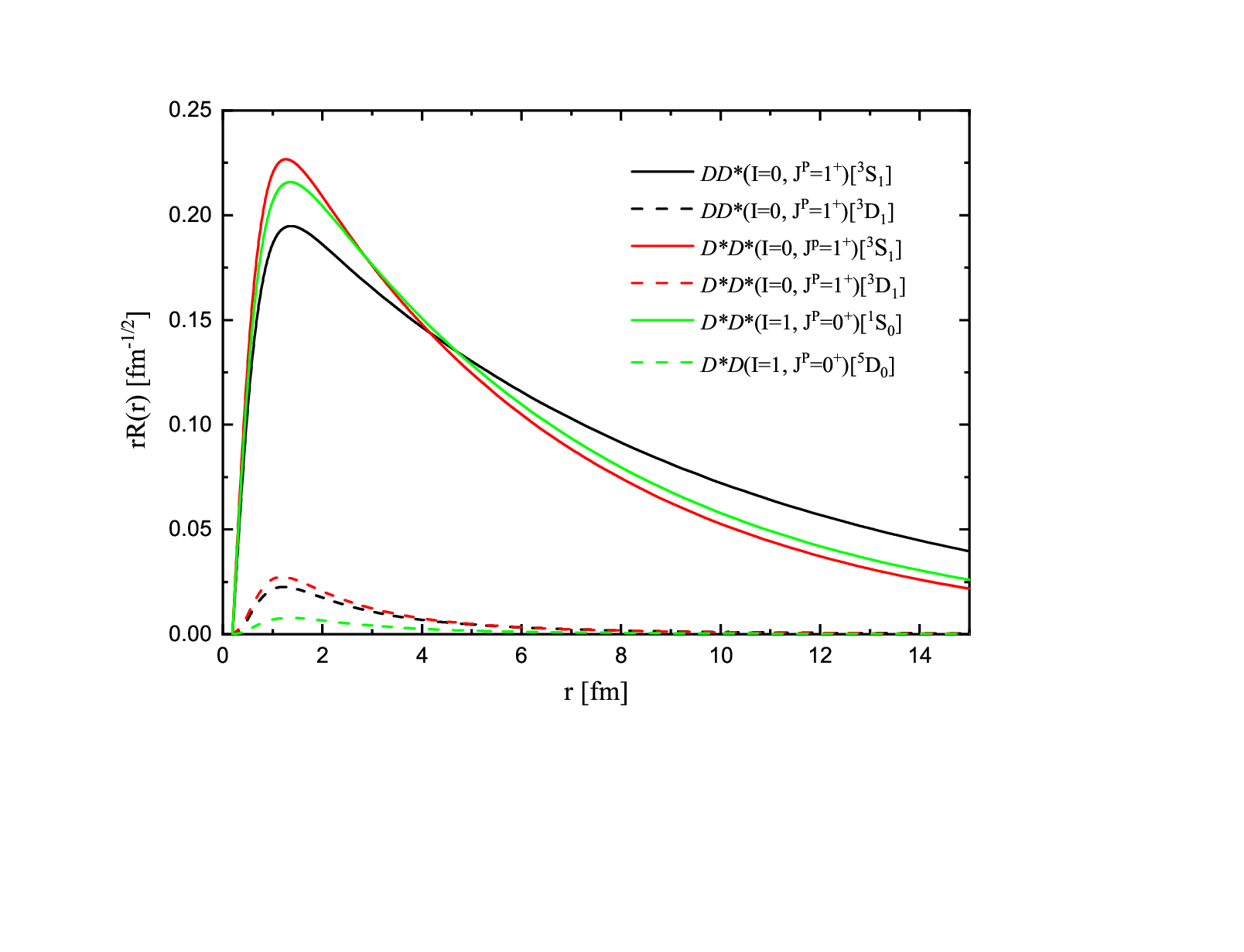}\\
\vspace*{-2cm}
\caption{(Color online) The radial wave functions of the bound states in the coordinate space for $D^{(\ast)}D^{(\ast)}$ system with $\Lambda$ =1.1426 GeV. }\label{waveDD}
\end{figure}

In Fig.\ref{waveDD}, we present the radial wave functions of the bound states in the coordinate space for $D^{(\ast)}D^{(\ast)}$ system. The black, red, and green lines represent the bound states $DD^{\ast}(I=0,J^P=1^+)$, $D^{\ast}D^{\ast}(I=0,J^P=1^+)$, and $D^{\ast}D^{\ast}(I=1,J^P=0^+)$  respectively, where the solid line represents the $S$-wave component and the dashed line represents the $D$-wave component. It can be seen that the contribution of all the bound states is mainly contributed by the $S$-wave component, while the contribution of the $D$-wave component can be ignored. The numerical results demonstrate that, the probabilities of the $S$-wave components are $99.53\%$, $99.35\%$ and $99.94\%$ for the bound states  $DD^{\ast}(I=0,J^P=1^+)$, $D^{\ast}D^{\ast}(I=0,J^P=1^+)$ and $D^{\ast}D^{\ast}(I=1,J^P=0^+)$ respectively, while the corresponding $D$-wave component is $0.47\%$, $0.65\%$ and $0.06\%$.
The RMS of the isospin vector bound state $D^{\ast}D^{\ast}(I=1,J^P=0^+)$ is 4.8 fm, and the RMS of the isospin scalar bound states $DD^{\ast}(I=0,J^P=1^+)$ and $D^{\ast}D^{\ast}(I=0,J^P=1^+)$ are 6.1 fm and 4.4 fm, as shown in Table~\ref{tbl:result_table_DD}.

\begin{figure}
\centering
\includegraphics[width=10cm]{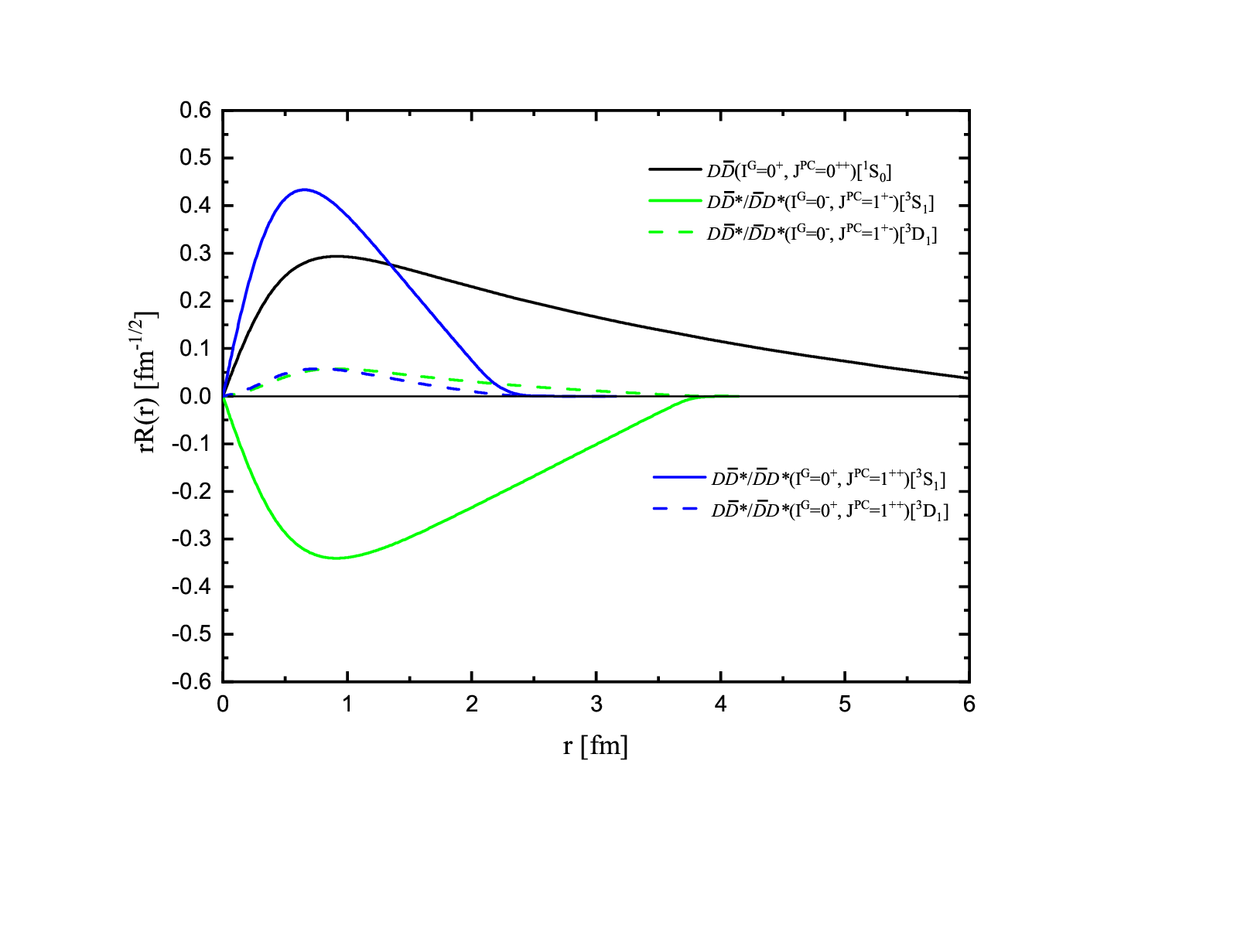}\\
\vspace*{-1.5cm}
\caption{(Color online) The radial wave functions of the bound states in the coordinate space for $D^{(\ast)}\bar{D}^{(\ast)}$ system with isospin $I=0$ at $\Lambda$ =1.073 GeV. }\label{waveDDbar2}
\end{figure}

\begin{figure}
\centering
\includegraphics[width=10cm]{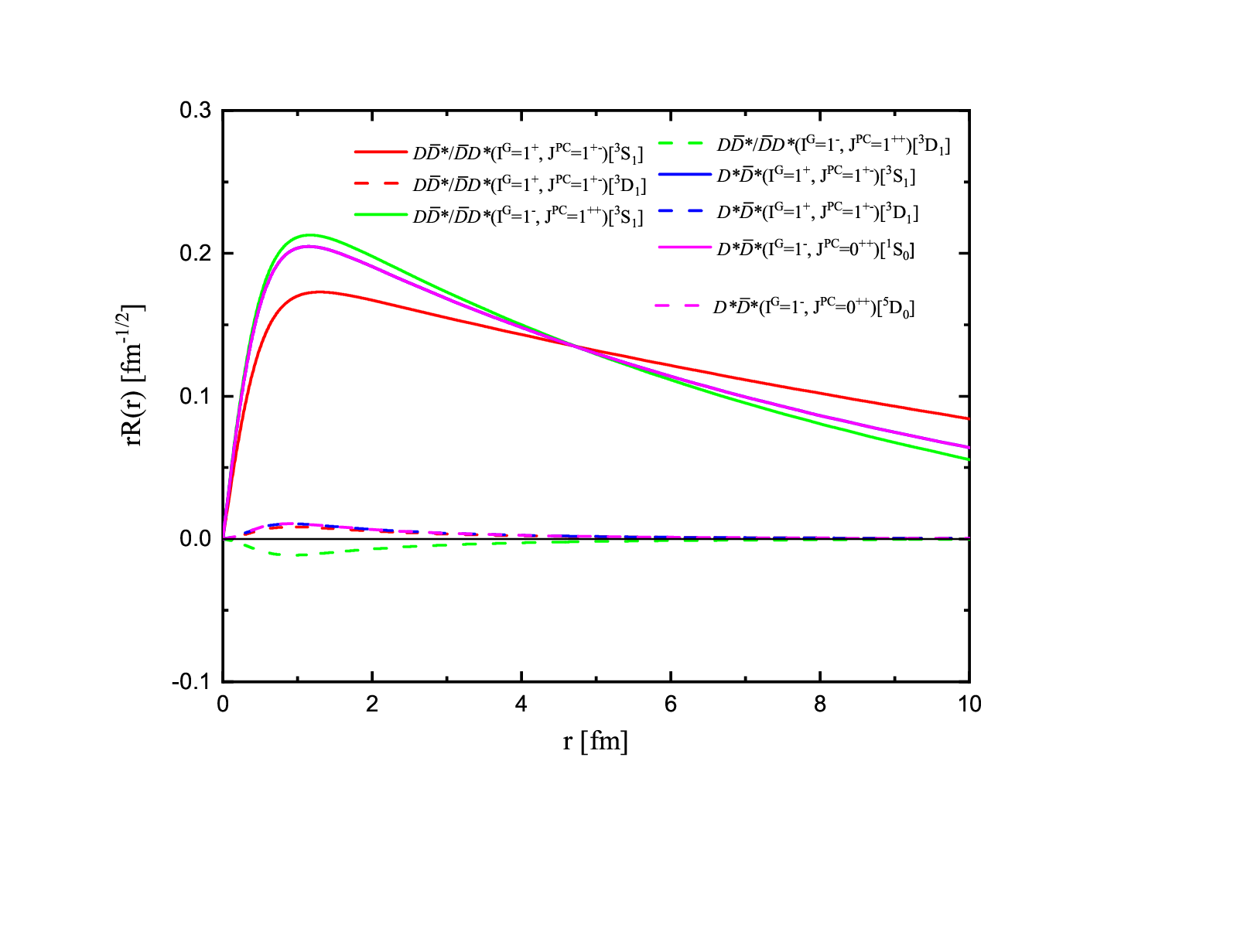}\\
\vspace*{-1.5cm}
\caption{(Color online) The radial wave functions of the bound states in the coordinate space for $D^{(\ast)}\bar{D}^{(\ast)}$ system with isospin $I=1$ at $\Lambda$ =1.25 GeV. }\label{waveDDbar}
\end{figure}

For $D^{(\ast)}\bar{D}^{(\ast)}$ system, the radial wave functions of the bound states in the coordinate space are presented in Fig.\ref{waveDDbar2} and Fig.\ref{waveDDbar}.
For $D^{(\ast)}\bar{D}^{(\ast)}$ system with isospin $I=0$ in Fig.\ref{waveDDbar2}, the numerical results demonstrate that, the probabilities of the $S$-wave components are $97.85\%$ and $98.44\%$ for the bound states  $D\bar{D}^{\ast}/\bar{D}D^{\ast}(I^G=0^-,J^{PC}=1^{+-})$ and $D\bar{D}^{\ast}/\bar{D}D^{\ast}(I^G=0^+,J^{PC}=1^{++})$ respectively, while the corresponding $D$-wave component are $2.15\%$ and $1.56\%$. For the bound state $D\bar{D}(I^G=0^+,J^{PC}=0^{++})$, the distribution of the radial wave function is diffuse, and the RMS radius reach 2.18 fm.
For $D^{(\ast)}\bar{D}^{(\ast)}$ system with isospin $I=1$ in Fig.\ref{waveDDbar}, the lines of wave functions for the states  $D^{\ast}\bar{D}^{\ast}(I^G=1^+,J^{PC}=1^{+-})$ and $D^{\ast}\bar{D}^{\ast}(I^G=1^-,J^{PC}=0^{++})$ almost overlap. Their probabilities of the $S$-wave components are all $99.8\%$, the contributions of the $D$-wave components can be ignored. The distributions of radial wave functions for the states $D\bar{D}^{\ast}/\bar{D}D^{\ast}(I^G=1^-,J^{PC}=1^{++})$ and $D^{\ast}\bar{D}^{\ast}(I^G=1^+,J^{PC}=1^{+-})$ are similar, but their peak positions move forward in sequence. The wave function of the $D\bar{D}(I^G=1^+,J^{PC}=1^{+-})$ state is the most dispersed in these states, and its RMS reaches 6.5 fm as shown in Table~\ref{tbl:result_table_DDbar}.

\section{Summary}\label{sec4}
In recent years, numerous exotic hadron states, such as the $X$, $Y$$, Z$ and $P_c$ particles, have been discovered experimentally. The exploration of these new hadron states structure and decay is a hot field in particle physics theory and experiments. In various theoretical models, hadron molecular state is a natural explanation for the exotic hadron states near two-hadron threshold.
In this work, we perform a systematic study of potential molecular states composed of heavy meson pairs, such as $D^{(*)}D^{(*)}$ and $D^{(*)}\bar{D}^{(*)}$, within the framework of the OBE model. The exchanged bosons include pseudoscalar, scalar, and vector mesons($\pi$, $\sigma$, $\rho$, $\omega$). Then, we solve the Schr\"{o}dinger equations for each quantum state of the $D^{(\ast)}D^{(\ast)}$ and $D^{(\ast)}\bar{D}^{(\ast)}$ systems with CSM in OBE potentials. The results indicated that the $D^{(*)}D^{(*)}$ and $D^{(*)}\bar{D}^{(*)}$ systems can form not only multiple bound states, but also several P-wave resonant states. In the OBE framework, the $X(3872)$, $T_{cc}^+$, and $Z_c(3900)$ states can be consistently explained as bound states, while the $G(3900)$ can be interpreted as a P-wave resonant state. Furthermore, we also predict other new bound and resonant states, which have the potential to be observed experimentally. We calculated the RMS of these bound states, and found that the radii of these states are between 2.7 fm and 6.1 fm, which are within the range of hadron molecular states. Furthermore, we calculated the proportion of different coupling channels and provided the wave functions of the bound states. The results indicate that the $S$-wave component contributed the main contribution, the contribution of the $D$-wave component can be ignored.

\vfill
\section*{ACKNOWLEDGMENTS}
This work was supported in part by the National Key R$\&$D Program of China (No.2022YFF0604801), the National Natural Science Foundation of China (No.11935001), the Natural Science Foundation of Anhui Province (No.2108085MA20, No.2208085MA10), and the Key Research Foundation of Education Ministry of Anhui Province of China (No.KJ2021A0061).

\appendix
\section{Hamiltonian Matrix}\label{appendix;H}
In this section, the kinetic terms and potential matrices for each $I(J^P)$ are given,
\subsection{kinetic energy matrix}\label{appendix;K}
For two pseudoscalar meson $DD$ system, the kinetic energy matrices for states $I(J^{P})$ are
\begin{align*}
  K_{0(1^{-})} &= \rm{diag} \left({-\frac{1}{2\mu_{DD}}\triangle_{1}}\right),\\
  K_{1(0^{+})} &= \rm{diag} \left({-\frac{1}{2\mu_{DD}}\triangle_{0}}\right),\\
  K_{1(2^{+})} &= \rm{diag} \left({-\frac{1}{2\mu_{DD}}\triangle_{2}}\right).\\
\end{align*}
For the pseudoscalar and meson vector $DD^{*}$ system, the kinetic energy matrices for states $I(J^{P})$ are
\begin{align*}
  K_{0(0^{-})} &= \rm{diag}\left({-\frac{1}{2\mu_{DD^{*}}}\triangle_{1}}\right),\\
  K_{0(1^{+})} &= \rm{diag}\left({-\frac{1}{2\mu_{DD^{*}}}\triangle_{0},-\frac{1}{2\mu_{DD^{*}}}\triangle_{2}}\right),\\
  K_{0(1^{-})} &= \rm{diag}\left({-\frac{1}{2\mu_{DD^{*}}}\triangle_{1}}\right),\\
  K_{0(2^{+})} &= \rm{diag}\left({-\frac{1}{2\mu_{DD^{*}}}\triangle_{2}}\right),\\
  K_{0(2^{-})} &= \rm{diag}\left({-\frac{1}{2\mu_{DD^{*}}}\triangle_{1},-\frac{1}{2\mu_{DD^{*}}}\triangle_{3}}\right),\\
  K_{1(0^{+})} &= \rm{diag}\left({-\frac{1}{2\mu_{DD^{*}}}\triangle_{0},-\frac{1}{2\mu_{DD^{*}}}\triangle_{2}}\right),\\
  K_{1(0^{-})} &= \rm{diag}\left({-\frac{1}{2\mu_{DD^{*}}}\triangle_{1}}\right),\\
  K_{1(1^{+})} &= \rm{diag}\left({-\frac{1}{2\mu_{DD^{*}}}\triangle_{0},-\frac{1}{2\mu_{DD^{*}}}\triangle_{2}}\right),\\
  K_{1(1^{-})} &= \rm{diag}\left({-\frac{1}{2\mu_{DD^{*}}}\triangle_{1}}\right),\\
  K_{1(2^{+})} &= \rm{diag}\left({-\frac{1}{2\mu_{DD^{*}}}\triangle_{2}}\right),\\
  K_{1(2^{-})} &= \rm{diag}\left({-\frac{1}{2\mu_{DD^{*}}}\triangle_{1},-\frac{1}{2\mu_{DD^{*}}}\triangle_{3}}\right),\\
\end{align*}
For two vector meson $D^{*}D^{*}$ system, the kinetic energy matrices for states $I(J^{P})$ are
\begin{align*}
  K_{0(1^{+})} &= \rm{diag}\left({-\frac{1}{2\mu_{D^{*}D^{*}}}\triangle_{0},-\frac{1}{2\mu_{D^{*}D^{*}}}\triangle_{2}}\right),\\
  K_{0(1^{-})} &= \rm{diag}\left({-\frac{1}{2\mu_{D^{*}D^{*}}}\triangle_{1},-\frac{1}{2\mu_{D^{*}D^{*}}}\triangle_{1},-\frac{1}{2\mu_{D^{*}D^{*}}}\triangle_{3}}\right),\\
  K_{0(2^{+})} &= \rm{diag}\left({-\frac{1}{2\mu_{D^{*}D^{*}}}\triangle_{2}}\right),\\
  K_{0(2^{-})} &= \rm{diag}\left({-\frac{1}{2\mu_{D^{*}D^{*}}}\triangle_{1},-\frac{1}{2\mu_{D^{*}D^{*}}}\triangle_{3}}\right),\\
  K_{1(1^{+})} &= \rm{diag}\left({-\frac{1}{2\mu_{D^{*}D^{*}}}\triangle_{2}}\right),\\
  K_{1(1^{-})} &= \rm{diag}\left({-\frac{1}{2\mu_{D^{*}D^{*}}}\triangle_{1}}\right),\\
  K_{1(2^{+})} &= \operatorname{diag}\left(-\frac{1}{2\mu_{P^{*}P^{*}}}\triangle_{2},-\frac{1}{2\mu_{P^{*}P^{*}}}\triangle_{0},\right.\\
  &\quad \left.-\frac{1}{2\mu_{P^{*}P^{*}}}\triangle_{2},-\frac{1}{2\mu_{P^{*}P^{*}}}\triangle_{4}\right),\\
  K_{1(2^{-})} &= \rm{diag}\left({-\frac{1}{2\mu_{D^{*}D^{*}}}\triangle_{1},-\frac{1}{2\mu_{D^{*}D^{*}}}\triangle_{3}}\right),\\
\end{align*}
where
\begin{align*}
  \mu_{D^{(*)}D^{(*)}} &= \frac{m_{D^{(*)}}m_{D^{(*)}}}{m_{D^{(*)}} + m_{D^{(*)}}},\\
  \triangle_{l} &= \frac{d^2}{dr^2} +\frac{2}{r}\frac{d}{dr} - \frac{l(l+1)}{r^2},\\
\end{align*}

\subsection{Potential matrix in the hadronic-molecule basis}\label{appendix;V_HM}
For $DD$ system, the potential matrices for states $I(J^{P})$ are
\begin{itemize}
  \item $0(1^{-})$
  \begin{align*}
    V_{v}^{0(1^{-})}=C^{\prime}_{v}
\end{align*}
\end{itemize}

\begin{itemize}
  \item $1(0^{+})$
  \begin{align*}
    V_{v}^{1(0^{+})}=C^{\prime}_{v}
\end{align*}
\end{itemize}

\begin{itemize}
  \item $1(2^{+})$
  \begin{align*}
    V_{v}^{1(2^{+})}=C^{\prime}_{v}
\end{align*}
\end{itemize}

For $DD^*$ system, the potential matrices for states $I(J^{P})$ are
\begin{itemize}
  \item $0(0^{-})$
  \begin{align*}
    V_{\pi}^{0(0^{-})}=C_{\pi} + 2 T_{\pi} \\
    V_{v}^{0(0^{-})}=2 C_{v} - 2 T_{v} + C^{\prime}_{v}
\end{align*}
\end{itemize}

\begin{itemize}
  \item $0(1^{+})$
  \begin{align*}
    V^{0(1^{+})}_{\pi}
    &=\begin{pmatrix}
      -C_{\pi}&\!\!\sqrt{2}T_{\pi}\\\vspace{1mm}
      \sqrt{2}T_{\pi}&\!\!-C_{\pi}-T_{\pi}
    \end{pmatrix},\\
\end{align*}

\begin{align*}
    V^{0(1^{+})}_{v}
    &=\begin{pmatrix}
      -2C_{v}+C^{\prime}_{v}&\!\!-\sqrt{2}T_{v}\\\vspace{1mm}
      -\sqrt{2}T_{v}&\!\!-2C_{v}+T_{v}+C^{\prime}_{v}
    \end{pmatrix},\\
\end{align*}
\end{itemize}

\begin{itemize}
  \item $0(1^{-})$
  \begin{align*}
    V^{0(1^{-})}_{\pi}= C_{\pi}- T_{\pi}\\
   V^{0(1^{-})}_{v}=2C_{v}+T_{v}+C^{\prime}_{v}
\end{align*}
\end{itemize}

\begin{itemize}
  \item $0(2^{+})$
  \begin{align*}
    V^{0(2^{+})}_{\pi}= -C_{\pi} + T_{\pi}\\
   V^{0(2^{+})}_{v}=-2C_{v}-T_{v}+C^{\prime}_{v}
\end{align*}
\end{itemize}

\begin{itemize}
  \item $0(2^{-})$
\begin{align*}
    V^{0(2^{-})}_{\pi}
    &=\begin{pmatrix}
      C_{\pi}+\frac{1}{5}T_{\pi}&\!\! -\frac{3\sqrt{6}}{5}T_{\pi}\\\vspace{1mm}
      -\frac{3\sqrt{6}}{5}T_{\pi}&\!\!C_{\pi}+\frac{4}{5}T_{\pi}
    \end{pmatrix},\\
\end{align*}

\begin{align*}
    V^{0(2^{-})}_{v}
    &=\begin{pmatrix}
      2C_{v}-\frac{1}{5}T_{v}+C^{\prime}_{v}&\!\! \frac{3\sqrt{6}}{5}T_{v}\\\vspace{1mm}
      \frac{3\sqrt{6}}{5}T_{v}&\!\!2C_{v}-\frac{4}{5}T_{v}+C^{\prime}_{v}
    \end{pmatrix},\\
\end{align*}
\end{itemize}

\begin{itemize}
  \item $1(0^{-})$
  \begin{align*}
    V^{1(0^{-})}_{\pi}= -C_{\pi}- 2T_{\pi}\\
   V^{1(0^{-})}_{v}=-2C_{v}+2T_{v}+C^{\prime}_{v}
\end{align*}
\end{itemize}

\begin{itemize}
  \item $1(1^{+})$
  \begin{align*}
    V^{1(1^{+})}_{\pi}
    &=\begin{pmatrix}
      C_{\pi}&\!\!-\sqrt{2}T_{\pi}\\\vspace{1mm}
      -\sqrt{2}T_{\pi}&\!\!C_{\pi}+T_{\pi}
    \end{pmatrix},\\
\end{align*}

\begin{align*}
    V^{1(1^{+})}_{v}
    &=\begin{pmatrix}
      2C_{v}+C^{\prime}_{v}&\!\!\sqrt{2}T_{v}\\\vspace{1mm}
      \sqrt{2}T_{v}&\!\!2C_{v}-T_{v}+C^{\prime}_{v}
    \end{pmatrix},\\
\end{align*}
\end{itemize}

\begin{itemize}
  \item $1(1^{-})$
  \begin{align*}
    V^{1(1^{-})}_{\pi}= -C_{\pi}+ T_{\pi}\\
   V^{1(1^{-})}_{v}=-2C_{v}-T_{v}+C^{\prime}_{v}
\end{align*}
\end{itemize}

\begin{itemize}
  \item $1(2^{+})$
  \begin{align*}
    V^{1(2^{+})}_{\pi}= C_{\pi} - T_{\pi}\\
   V^{1(2^{+})}_{v}=2C_{v}+T_{v}+C^{\prime}_{v}
\end{align*}

\begin{align*}
    V^{1(2^{-})}_{\pi}
    &=\begin{pmatrix}
      -C_{\pi}-\frac{1}{5}T_{\pi}&\!\! \frac{3\sqrt{6}}{5}T_{\pi}\\\vspace{1mm}
      \frac{3\sqrt{6}}{5}T_{\pi}&\!\!-C_{\pi}-\frac{4}{5}T_{\pi}
    \end{pmatrix},\\
\end{align*}

\begin{align*}
    V^{1(2^{-})}_{v}
    &=\begin{pmatrix}
      -2C_{v}+\frac{1}{5}T_{v}+C^{\prime}_{v}&\!\! -\frac{3\sqrt{6}}{5}T_{v}\\\vspace{1mm}
      -\frac{3\sqrt{6}}{5}T_{v}&\!\!-2C_{v}+\frac{4}{5}T_{v}+C^{\prime}_{v}
    \end{pmatrix},\\
\end{align*}
\end{itemize}

For $D^*D^*$ system, the potential matrices for states $I(J^{P})$ are
\begin{itemize}
  \item $0(1^{+})$
  \begin{align*}
    V^{0(1^{+})}_{\pi}
    &=\begin{pmatrix}
      -C_{\pi}&\!\!\sqrt{2}T_{\pi}\\\vspace{1mm}
      \sqrt{2}T_{\pi}&\!\!-C_{\pi}-T_{\pi}
    \end{pmatrix},\\
\end{align*}
\begin{align*}
    V^{0(1^{+})}_{v}
    &=\begin{pmatrix}
      -2C_{v}+C^{\prime}_{v}&\!\!-\sqrt{2}T_{v}\\\vspace{1mm}
      -\sqrt{2}T_{v}&\!\!-2C_{v}+T_{v}+C^{\prime}_{v}
    \end{pmatrix},\\
\end{align*}
\end{itemize}

\begin{itemize}
  \item $0(1^{-})$
  \begin{align*}
    V^{0(1^{-})}_{\pi}
    &=\begin{pmatrix}
      -2C_{\pi}&\!\!\frac{2}{\sqrt{5}}T_{\pi}&\!\!-\sqrt{\frac{6}{5}}T_{\pi}\\\vspace{1mm}
      \frac{2}{\sqrt{5}}T_{\pi}&\!\!C_{\pi}-\frac{7}{5}T_{\pi}&\!\!\frac{\sqrt{6}}{5}T_{\pi}\\\vspace{1mm}
      -\sqrt{\frac{6}{5}}T_{\pi}&\!\!\frac{\sqrt{6}}{5}T_{\pi}&\!\!C_{\pi}-\frac{8}{5}T_{\pi}
    \end{pmatrix},\\
\end{align*}

\begin{align*}
    V^{0(1^{-})}_{v}
    &=\begin{pmatrix}
      -4C_{v}+C^{\prime}_{v}&\!\!-\frac{2}{\sqrt{5}}T_{v}&\!\!\sqrt{\frac{6}{5}}T_{v}\\\vspace{1mm}
      -\frac{2}{\sqrt{5}}T_{v}&\!\!2C_{v}+\frac{7}{5}T_{v}+C^{\prime}_{v}&\!\!-\frac{\sqrt{6}}{5}T_{v}\\\vspace{1mm}
      \sqrt{\frac{6}{5}}T_{v}&\!\!-\frac{\sqrt{6}}{5}T_{v}&\!\!2C_{v}+\frac{8}{5}T_{v}+C^{\prime}_{v}
    \end{pmatrix},\\
\end{align*}
\end{itemize}

\begin{itemize}
  \item $0(2^{+})$
  \begin{align*}
    V^{0(2^{+})}_{\pi}= -C_{\pi} + T_{\pi}\\
   V^{0(2^{+})}_{v}=-2C_{v}-T_{v}+C^{\prime}_{v}
\end{align*}
\end{itemize}

\begin{itemize}
  \item $0(2^{-})$
\begin{align*}
    V^{0(2^{-})}_{\pi}
    &=\begin{pmatrix}
      C_{\pi}+\frac{7}{5}T_{\pi}&\!\! -\frac{3\sqrt{6}}{5}T_{\pi}\\\vspace{1mm}
      -\frac{3\sqrt{6}}{5}T_{\pi}&\!\!C_{\pi}-\frac{2}{5}T_{\pi}
    \end{pmatrix},\\
\end{align*}

\begin{align*}
    V^{0(2^{-})}_{v}
    &=\begin{pmatrix}
      2C_{v}-\frac{7}{5}T_{v}+C^{\prime}_{v}&\!\! -\frac{6}{5}T_{v}\\\vspace{1mm}
      -\frac{6}{5}T_{v}&\!\!2C_{v}+\frac{2}{5}T_{v}+C^{\prime}_{v}
    \end{pmatrix},\\
\end{align*}
\end{itemize}

\begin{itemize}
  \item $1(0^{+})$
\begin{align*}
    V^{1(0^{+})}_{\pi}
    &=\begin{pmatrix}
      -2C_{\pi}&\!\! -\sqrt{2}T_{\pi}\\\vspace{1mm}
      -\sqrt{2}T_{\pi}&\!\!C_{\pi}-2T_{\pi}
    \end{pmatrix},\\
\end{align*}

\begin{align*}
    V^{1(0^{+})}_{v}
    &=\begin{pmatrix}
      -4C_{v}+C^{\prime}_{v}&\!\! \sqrt{2}T_{v}\\\vspace{1mm}
      \sqrt{2}T_{v}&\!\!2C_{v}+2T_{v}+C^{\prime}_{v}
    \end{pmatrix},\\
\end{align*}
\end{itemize}

\begin{itemize}
  \item $1(0^{-})$
  \begin{align*}
    V^{1(0^{-})}_{\pi}= -C_{\pi}- 2T_{\pi}\\
   V^{1(0^{-})}_{v}=-2C_{v}+2T_{v}+C^{\prime}_{v}
\end{align*}
\end{itemize}

\begin{itemize}
  \item $1(1^{+})$
  \begin{align*}
    V^{1(1^{+})}_{\pi}= C_{\pi} - T_{\pi}\\
   V^{1(1^{+})}_{v}=2C_{v}+T_{v}+C^{\prime}_{v}
\end{align*}

\end{itemize}

\begin{itemize}
  \item $1(1^{-})$
  \begin{align*}
    V^{1(1^{-})}_{\pi}= -C_{\pi}+ T_{\pi}\\
   V^{1(1^{-})}_{v}=-2C_{v}-T_{v}+C^{\prime}_{v}
\end{align*}
\end{itemize}

\begin{itemize}
  \item $1(2^{+})$
  \begin{align*}
    V^{1(2^{+})}_{\pi}
    &=\begin{pmatrix}
      -2C_{\pi}&\!\!-\sqrt{\frac{2}{5}}T_{\pi}&\!\!\frac{2}{\sqrt{7}}T_{\pi}&\!\!-\frac{6}{\sqrt{35}}T_{\pi}\\\vspace{1mm}
      -\sqrt{\frac{2}{5}}T_{\pi}&\!\!C_{\pi}&\!\!\sqrt{\frac{14}{5}}T_{\pi}&\!\!0\\\vspace{1mm}
      \frac{2}{\sqrt{7}}T_{\pi}&\!\!\sqrt{\frac{14}{5}}T_{\pi}&\!\!C_{\pi}+\frac{3}{7}T_{\pi}&\!\!\frac{12}{7\sqrt{5}}T_{\pi}\\\vspace{1mm}
      -\frac{6}{\sqrt{35}}T_{\pi}&\!\!0&\!\!\frac{12}{7\sqrt{5}}T_{\pi}&\!\!C_{\pi}-\frac{10}{7}T_{\pi}
    \end{pmatrix},\\
\end{align*}

\begin{align*}
    V^{1(2^{+})}_{v}
    &=\begin{pmatrix}
      -4C_{v}+C^{\prime}_{v}&\!\!\sqrt{\frac{2}{5}}T_{v}&\!\!-\frac{2}{\sqrt{7}}T_{v}&\!\!\frac{6}{\sqrt{35}}T_{v}\\\vspace{1mm}
     \sqrt{\frac{2}{5}}T_{v}&\!\!C_{v}+C^{\prime}_{v}&\!\!-\sqrt{\frac{14}{5}}T_{v}&\!\!0\\\vspace{1mm}
      -\frac{2}{\sqrt{7}}T_{v}&\!\!-\sqrt{\frac{14}{5}}T_{v}&\!\!2C_{\pi}-\frac{3}{7}T_{v}+C^{\prime}_{v}&\!\!-\frac{12}{7\sqrt{5}}T_{v}\\\vspace{1mm}
      \frac{6}{\sqrt{35}}T_{\pi}&\!\!0&\!\!-\frac{12}{7\sqrt{5}}T_{\pi}&\!\!2C_{\pi}+\frac{10}{7}T_{\pi}+C^{\prime}_{v}
    \end{pmatrix},\\
\end{align*}
\end{itemize}

\begin{itemize}
\item $1(2^{-})$
\begin{align*}
    V^{1(2^{-})}_{\pi}
    &=\begin{pmatrix}
      -C_{\pi}-\frac{1}{5}T_{\pi}&\!\! \frac{3\sqrt{6}}{5}T_{\pi}\\\vspace{1mm}
      \frac{3\sqrt{6}}{5}T_{\pi}&\!\!-C_{\pi}-\frac{4}{5}T_{\pi}
    \end{pmatrix},\\
\end{align*}

\begin{align*}
    V^{1(2^{-})}_{v}
    &=\begin{pmatrix}
      -2C_{v}+\frac{1}{5}T_{v}+C^{\prime}_{v}&\!\! -\frac{3\sqrt{6}}{5}T_{v}\\\vspace{1mm}
      -\frac{3\sqrt{6}}{5}T_{v}&\!\!-2C_{v}+\frac{4}{5}T_{v}+C^{\prime}_{v}
    \end{pmatrix},\\
\end{align*}
\end{itemize}
The $\sigma$ exchange has been taken into account, which leads to diagonal interaction for all possible states. The form of matrix elements is as follows,

\begin{align*}
    V_{\sigma} = C_{\sigma},
\end{align*}
\begin{align*}
    V_{\sigma}
    &=\begin{pmatrix}
      C_{\sigma}&\!\!0 \\\vspace{1mm}
      0&\!\!C_{\sigma}
    \end{pmatrix},\\
\end{align*}
\begin{align*}
    V_{\sigma}
    &=\begin{pmatrix}
      C_{\sigma}&\!\! 0 &\!\! 0 \\\vspace{1mm}
      0 &\!\! C_{\sigma} &\!\! 0 \\\vspace{1mm}
      0 &\!\! 0 &\!\! C_{\sigma}
    \end{pmatrix}.\\
\end{align*}

In matrix elements, $C_{\pi}$, $T_{\pi}$, $C_{v}$, $T_{v}$, $C^{\prime}_{v}$ and $C_{\sigma}$ are defined as

\begin{align*}
    C_{\pi} &= \frac{1}{3}{\left(\frac{g}{2f_{\pi}}\right)}^2 C(r;m_{\pi})\vec{\tau}_{1}\cdot\vec{\tau}_{2},\\
    T_{\pi} &= \frac{1}{3}{\left(\frac{g}{2f_{\pi}}\right)}^2 T(r;m_{\pi})\vec{\tau}_{1}\cdot\vec{\tau}_{2},\\
    C_{v} &= \frac{1}{3}{(\lambda g_{V})}^2 C(r;m_{v})\vec{\tau}_{1}\cdot\vec{\tau}_{2},\\
    T_{v} &= \frac{1}{3}{(\lambda g_{V})}^2 T(r;m_{v})\vec{\tau}_{1}\cdot\vec{\tau}_{2},\\
    C^{\prime}_{v} &= {\left(\frac{\beta g_{V}}{2m_{v}}\right)}^2 C(r;m_{v})\vec{\tau}_{1}\cdot\vec{\tau}_{2},\\
    C_{\sigma} &= -{\left(\frac{g_s}{m_{\sigma}}\right)}^2 C(r;m_{\sigma}).
\end{align*}


\begin{thebibliography}{99}
\bibitem{Gell-Mann:1964}
M. ~Gell-Mann,
A schematic model of baryons and mesons,
Physics Letters,
\textbf{8}, 3, 214-215 (1964).


\bibitem{Belle:2003nnu}
S.~K.~Choi \textit{et al.} [Belle],
Phys. Rev. Lett. \textbf{91}, 262001 (2003).

\bibitem{CDF:2003cab}
D.~Acosta \textit{et al.} [CDF],
Phys. Rev. Lett. \textbf{93}, 072001 (2004).

\bibitem{D0:2004zmu}
V.~M.~Abazov \textit{et al.} [D0],
Phys. Rev. Lett. \textbf{93}, 162002 (2004).

\bibitem{BaBar:2004oro}
B.~Aubert \textit{et al.} [BaBar],
Phys. Rev. D \textbf{71}, 071103 (2005).

\bibitem{Xu:2017tsr}
H.~Xu, B.~Wang, Z.~W.~Liu and X.~Liu,
Phys. Rev. D \textbf{99}, 014027 (2019)
[erratum: Phys. Rev. D \textbf{104}, 119903 (2021)].

\bibitem{Ohkoda:2012hv}
S.~Ohkoda, Y.~Yamaguchi, S.~Yasui, K.~Sudoh and A.~Hosaka,
Phys. Rev. D \textbf{86}, 034019 (2012).

\bibitem{Li:2012cs}
N.~Li and S.~L.~Zhu,
Phys. Rev. D \textbf{86}, 074022 (2012).

\bibitem{Ren:2021dsi}
H.~Ren, F.~Wu and R.~Zhu,
Adv. High Energy Phys. \textbf{2022}, 9103031 (2022).


\bibitem{Sakai:2017avl}
S.~Sakai, L.~Roca and E.~Oset,
Phys. Rev. D \textbf{96}, 054023 (2017).

\bibitem{He:2014nya}
J.~He,
Phys. Rev. D \textbf{90}, 076008 (2014).

\bibitem{He:2015mja}
J.~He,
Phys. Rev. D \textbf{92}, 034004 (2015).

\bibitem{Wallbott:2019dng}
P.~C.~Wallbott, G.~Eichmann and C.~S.~Fischer,
Phys. Rev. D \textbf{100}, 014033 (2019).

\bibitem{Zhu:2019iwm}
R.~Zhu, X.~Liu, H.~Huang and C.~F.~Qiao,
Phys. Lett. B \textbf{797}, 134869 (2019).

\bibitem{Ortega:2021yis}
P.~G.~Ortega, D.~R.~Entem and F.~Fern\'andez,
Phys. Lett. B \textbf{829}, 137083 (2022).

\bibitem{Tan:2020ldi}
Y.~Tan, W.~Lu and J.~Ping,
Eur. Phys. J. Plus \textbf{135}, 716 (2020).

\bibitem{Luo:2017eub}
S.~Q.~Luo, K.~Chen, X.~Liu, Y.~R.~Liu and S.~L.~Zhu,
Eur. Phys. J. C \textbf{77}, 709 (2017).

\bibitem{Navarra:2007yw}
F.~S.~Navarra, M.~Nielsen and S.~H.~Lee,
Phys. Lett. B \textbf{649}, 166-172 (2007).

\bibitem{Xin:2021wcr}
Q.~Xin and Z.~G.~Wang,
Eur. Phys. J. A \textbf{58}, 110 (2022).

\bibitem{Tang:2019nwv}
L.~Tang, B.~D.~Wan, K.~Maltman and C.~F.~Qiao,
Phys. Rev. D \textbf{101}, 094032 (2020).

\bibitem{Lu:2020rog}
Q.~F.~L\"u, D.~Y.~Chen and Y.~B.~Dong,
Phys. Rev. D \textbf{102}, 034012 (2020).

\bibitem{Ebert:2007rn}
D.~Ebert, R.~N.~Faustov, V.~O.~Galkin and W.~Lucha,
Phys. Rev. D \textbf{76}, 114015 (2007).

\bibitem{Wang:2018pwi}
G.~J.~Wang, X.~H.~Liu, L.~Ma, X.~Liu, X.~L.~Chen, W.~Z.~Deng and S.~L.~Zhu,
Eur. Phys. J. C \textbf{79}, 567 (2019).


\bibitem{Wigner:1947zz}
E.~P.~Wigner and L.~Eisenbud,
Phys. Rev. \textbf{72} (1947), 29-41.

\bibitem{Hale:1987zz}
G.~M.~Hale, R.~E.~Brown and N.~Jarmie,
Phys. Rev. Lett. \textbf{59} (1987), 763-766.

\bibitem{Humblet:1991zz}
J.~Humblet, B.~W.~Filippone and S.~E.~Koonin,
Phys. Rev. C \textbf{44} (1991), 2530-2535

\bibitem{Taylor}
J. R. Taylor, Scattering Theory: The Quantum Theory on
Nonrelativistic Collisions (John Wiley $\&$ Sons, New York, 1972).

\bibitem{Hazi}
A. U. Hazi and H. S. Taylor,
Phys. Rev. A \textbf{1}, 1109 (1970).

\bibitem{Kukulin}
V. I. Kukulin, V. M. Krasnoplsky, and J. Horacek,
Theory of Resonances: Principles and Applications (Kluwer, Dordrecht, The Netherlands, 1989).

\bibitem{csm1}
Y. K. Ho,
Phys. Rep. \textbf{99}, 1 (1983).

\bibitem{csm2}
N. Moiseyev,
Phys. Rep. \textbf{302}, 212 (1998).


\bibitem{Yu:2021lmb}
Z.~Yu, M.~Song, J.~Y.~Guo, Y.~Zhang and G.~Li,
Phys. Rev. C \textbf{104}, 035201 (2021).

\bibitem{Wang:2022yes}
G.~J.~Wang, Q.~Meng and M.~Oka,
Phys. Rev. D \textbf{106}, 096005 (2022).

\bibitem{Cheng:2022qcm}
J.~B.~Cheng, Z.~Y.~Lin and S.~L.~Zhu,
Phys. Rev. D \textbf{106}, 016012 (2022).

\bibitem{Wang:2023ivd}
Z.~P.~Wang, F.~L.~Wang, G.~J.~Wang and X.~Liu,
Phys. Rev. D \textbf{110}, L051501 (2024).

\bibitem{Lin:2024prl}
Z.~Y.~Lin, J.~Z.~Wang, J.~B.~Cheng, L.~Meng and S.~L.~Zhu,
Phys. Rev. Lett. \textbf{133}, 241903 (2024).

\bibitem{Lin:2022wmj}
Z.~Y.~Lin, J.~B.~Cheng and S.~L.~Zhu,
Phys. Rev. D \textbf{110}, 5 (2024).

\bibitem{Chen:2023eri}
Y.~K.~Chen, L.~Meng, Z.~Y.~Lin and S.~L.~Zhu,
Phys. Rev. D \textbf{109}, 034006 (2024).

\bibitem{Li:2012ss}
N.~Li, Z.~F.~Sun, X.~Liu and S.~L.~Zhu,
Phys. Rev. D \textbf{88} 114008, (2013).

\bibitem{Liu:2019stu}
M.~Z.~Liu, T.~W.~Wu, M.~Pavon Valderrama, J.~J.~Xie and L.~S.~Geng,
Phys. Rev. D \textbf{99}, 094018 (2019).

\bibitem{Abreu:2022sra}
L.~M.~Abreu,
Nucl. Phys. B \textbf{985}, 115994 (2022).

\bibitem{Abreu:2015jma}
L.~M.~Abreu,
Nucl. Phys. A \textbf{940}, 1-20 (2015).

\bibitem{Whyte:2024ihh}
T.~Whyte \textit{et al.} [Hadron Spectrum],
Phys. Rev. D \textbf{111} 034511, (2025).


\bibitem{Wang:2013kva}
P.~Wang and X.~G.~Wang,
Phys. Rev. Lett. \textbf{111}, 042002 (2013).

\bibitem{Thomas:2008ja}
C.~E.~Thomas and F.~E.~Close,
Phys. Rev. D \textbf{78}, 034007 (2008).

\bibitem{Braaten:2010mg}
E.~Braaten, H.~W.~Hammer and T.~Mehen,
Phys. Rev. D \textbf{82}, 034018 (2010).


\bibitem{Belle:2005lfc}
K.~Abe \textit{et al.} [Belle],
[arXiv:hep-ex/0505037 [hep-ex]].

\bibitem{BaBar:2010wfc}
P.~del Amo Sanchez \textit{et al.} [BaBar],
Phys. Rev. D \textbf{82} (2010), 011101.


\bibitem{Sakai:2023syt}
M.~Sakai and Y.~Yamaguchi,
Phys. Rev. D \textbf{109}, 054016 (2024).

\bibitem{Guo:2013sya}
F.~K.~Guo, C.~Hidalgo-Duque, J.~Nieves and M.~PavonValderrama,
Phys. Rev. D \textbf{88} (2013), 054007.

\bibitem{Wang:2013cya}
Q.~Wang, C.~Hanhart and Q.~Zhao,
Phys. Rev. Lett. \textbf{111}, 132003 (2013).

\bibitem{Huang:2025rvj}
Y.~Huang and X.~Chen,
[arXiv:2501.10992 [hep-ph]].

\bibitem{Chen:2025gxe}
X.~X.~Chen, Z.~M.~Ding and J.~He,
Phys. Rev. D \textbf{111}, 11 (2025).

\bibitem{Liu:2025sjz}
S.~D.~Liu, Q.~Wu and G.~Li,
[arXiv:2506.18273 [hep-ph]].

\bibitem{Ye:2025ywy}
Q.~Ye, Z.~Zhang, M.~L.~Du, U.~G.~Mei{\ss}ner, P.~Y.~Niu and Q.~Wang,
Phys. Rev. D \textbf{112}, 016015 (2025).

\bibitem{Nambu:1961tp}
Y.~Nambu and G.~Jona-Lasinio,
Phys.\ Rev.\  {\bf 122}, 345 (1961); Phys.\ Rev.\  {\bf 124}, 246 (1961).

\bibitem{Burdman:1992gh}
  G.~Burdman and J.~F.~Donoghue,
  Phys.\ Lett.\ B {\bf 280}, 287 (1992).

\bibitem{Wise:1992hn}
  M.~B.~Wise,
  Phys.\ Rev.\  D {\bf 45}, R2188 (1992).

\bibitem{Yan:1992gz}
  T.~M.~Yan, H.~Y.~Cheng, C.~Y.~Cheung, G.~L.~Lin, Y.~C.~Lin and H.~L.~Yu,
  Phys.\ Rev.\  D {\bf 46}, 1148 (1992)
  [Erratum-ibid.\  D {\bf 55}, 5851 (1997)].


\bibitem{Casalbuoni:1996pg}
  R.~Casalbuoni, A.~Deandrea, N.~Di Bartolomeo, R.~Gatto, F.~Feruglio and G.~Nardulli,
  Phys.\ Rept.\  {\bf 281}, 145 (1997).

\bibitem{Manohar:2000dt}
  A.~V.~Manohar and M.~B.~Wise,
  Camb.\ Monogr.\ Part.\ Phys.\ Nucl.\ Phys.\ Cosmol.\  {\bf 10}, 1 (2000).

\bibitem{Isola:2003fh}
  C.~Isola, M.~Ladisa, G.~Nardulli and P.~Santorelli,
  Phys.\ Rev.\  D {\bf 68}, 114001 (2003).

\bibitem{Ding:2008gr}
G.~J.~Ding,
Phys. Rev. D \textbf{79} (2009), 014001.

\bibitem{Wang:2022mxy}
F.~L.~Wang and X.~Liu,
Phys. Lett. B \textbf{835}, 137583 (2022).

\bibitem{Liu:2009qhy}
X.~Liu, Z.~G.~Luo, Y.~R.~Liu and S.~L.~Zhu,
Eur. Phys. J. C \textbf{61}, 411-428 (2009).

\bibitem{Wang:2020dya}
F.~L.~Wang and X.~Liu,
Phys. Rev. D \textbf{102}, 094006 (2020).

\bibitem{Chen:2021tip}
R.~Chen,
Eur. Phys. J. C \textbf{81}, 122 (2021).

\bibitem{Wang:2024ukc}
J.~Z.~Wang, Z.~Y.~Lin, B.~Wang, L.~Meng and S.~L.~Zhu,
Phys. Rev. D \textbf{110}, 114003 (2024).

\bibitem{Liu:2019zvb}
M.~Z.~Liu, T.~W.~Wu, M.~S{\'a}nchez S{\'a}nchez, M.~P.~Valderrama, L.~S.~Geng and J.~J.~Xie,
Phys. Rev. D \textbf{103}, 054004 (2021).

\bibitem{Ling:2021asz}
X.~Z.~Ling, M.~Z.~Liu and L.~S.~Geng,
Eur. Phys. J. C \textbf{81}, 1090 (2021).

\bibitem{Xu:2025mhc}
R.~Xu, L.~Meng, H.~X.~Zhu, N.~Li and W.~Chen,
Phys. Rev. D \textbf{111}, 094015 (2025).

\bibitem{Yalikun:2025ssz}
N.~Yalikun, X.~K.~Dong and U.~G.~Mei\ss{}ner,
[arXiv:2503.01322 [hep-ph]].


\bibitem{Ohkoda:2011vj}
  S.~Ohkoda, Y.~Yamaguchi, S.~Yasui, K.~Sudoh and A.~Hosaka,
  Phys. Rev. D \textbf{86}, 014004 (2012).

\bibitem{Yasui:2009bz}
  S.~Yasui and K.~Sudoh,
  Phys.\ Rev.\  D {\bf 80}, 034008 (2009).

\bibitem{Yamaguchi:2011xb}
  Y.~Yamaguchi, S.~Ohkoda, S.~Yasui and A.~Hosaka,
  Phys.\ Rev.\  D {\bf 84}, 014032 (2011).

\bibitem{abc}
J. Aguilar and J. M. Combes, Commun. Math. Phys. \textbf{22}, 269
(1971); E. Balslev and J. M. Combes, ibid. \textbf{22}, 280 (1971).

\bibitem{Ke:2021rxd}
H.~W.~Ke, X.~H.~Liu and X.~Q.~Li,
Eur. Phys. J. C \textbf{82}, 144 (2022).

\bibitem{LHCb:2021vvq}
R.~Aaij \textit{et al.} [LHCb],
Nature Phys. \textbf{18}, no.7, 751-754 (2022).

\bibitem{LHCb:2021auc}
R.~Aaij \textit{et al.} [LHCb],
Nature Commun. \textbf{13}, no.1, 3351 (2022).

\bibitem{Du:2021zzh}
M.~L.~Du, V.~Baru, X.~K.~Dong, A.~Filin, F.~K.~Guo, C.~Hanhart, A.~Nefediev, J.~Nieves and Q.~Wang,
Phys. Rev. D \textbf{105}, 014024 (2022).

\bibitem{Liu:2025fhl}
Z.~Liu, H.~Xu and X.~Liu,
[arXiv:2503.10299 [hep-ph]].

\bibitem{BESIII:2024ths}
M.~Ablikim \textit{et al.} [BESIII],
Phys. Rev. Lett. \textbf{133}, 081901 (2024).

\bibitem{Meng:2021jnw}
L.~Meng, G.~J.~Wang, B.~Wang and S.~L.~Zhu,
Phys. Rev. D \textbf{104}, L051502 (2021).


\end{thebibliography}
\end{document}